\documentclass[footinbib,aps,prl,twocolumn,showpacs,amsfonts,amssymb,floatfix]{revtex4-1}
\usepackage[T1]{fontenc}
\usepackage[utf8]{inputenc}
\usepackage{float}
\usepackage{graphicx}
\usepackage{amssymb}
\usepackage{amsmath}
\usepackage{enumerate}

\makeatletter

\newcommand*{\diff}{\mathop{\!\mathrm{d}\!}}
\date{February 8, 2020} 

\makeatother

\begin{document}

\title{ Dry Active Matter exhibits a self-organized ``Cross Sea'' Phase}

\author{Rüdiger Kürsten}
\affiliation{Institut für Physik, Universität Greifswald, Felix-Hausdorff-Str. 6, 17489 Greifswald, Germany}
\author{Thomas Ihle}
\affiliation{Institut für Physik, Universität Greifswald, Felix-Hausdorff-Str. 6, 17489 Greifswald, Germany}


\begin{abstract}
The Vicsek model of self-propelled particles is known in three different phases: $(i)$ a polar ordered homogeneous phase also called Toner-Tu phase, $(iii)$ a phase of polar ordered regularly arranged high density bands (waves) with surrounding low density regions without polar order and $(iv)$ a homogeneous phase without polar order.
It has been questioned whether the band phase $(iii)$ should be divided into two parts \cite{Chate20}: one with periodically arranged and one with strongly interacting but not ordered bands.
We answer this question by showing that the standard Vicsek model has a fourth phase for large system sizes: $(ii)$ a polar ordered cross sea phase.
Close to the transition towards $(i)$ this phase becomes unstable and looks like strongly interacting bands.
We demonstrate that the cross sea phase is not just a superposition of two waves, but it is an independent complex pattern.
Furthermore we show that there is a non-zero mass flow through the structure of the cross sea pattern within its co-moving frame.
\end{abstract}
\maketitle

Active matter is characterized by the transformation of free energy into directed motion.
The energy is supplied e.g. by chemicals (or food), external fields or radiation.
On the other hand, active particles dissipate energy into their environment such that there is an interplay between energy supply and dissipation.
Such processes are clearly far from thermodynamic equilibrium.

Active entities appear from micro length scales or even below e.g. for bacteria, Janus particles or molecular motors up to macroscopic sizes such as for birds, fish, mammals or robots.
They might be living organism or artificially manufactured non-living objects. 
Several reviews give an overview of the field \cite{Chate20, VZ12, RBELS12, Ramaswamy10, TTR05}.
Theoretical descriptions involve field- or kinetic theories, see e.g. \cite{TT95, TT98, BDG06, BDG09, Toner12, Ihle11, Ihle16, NI19}.

Usually, active particles are surrounded by a fluid like water or air. 
In many cases the fluid is important, in particular due to the conservation of momentum.
Examples are swimming bacteria or artificial microswimmers that are subject to intense research over the last decades, see e.g. \cite{SAHS03, PSJSDB12, SA12, DHDWBG13, GRLC14, KPV19}.

\begin{figure}
	\includegraphics[width=0.22\textwidth]{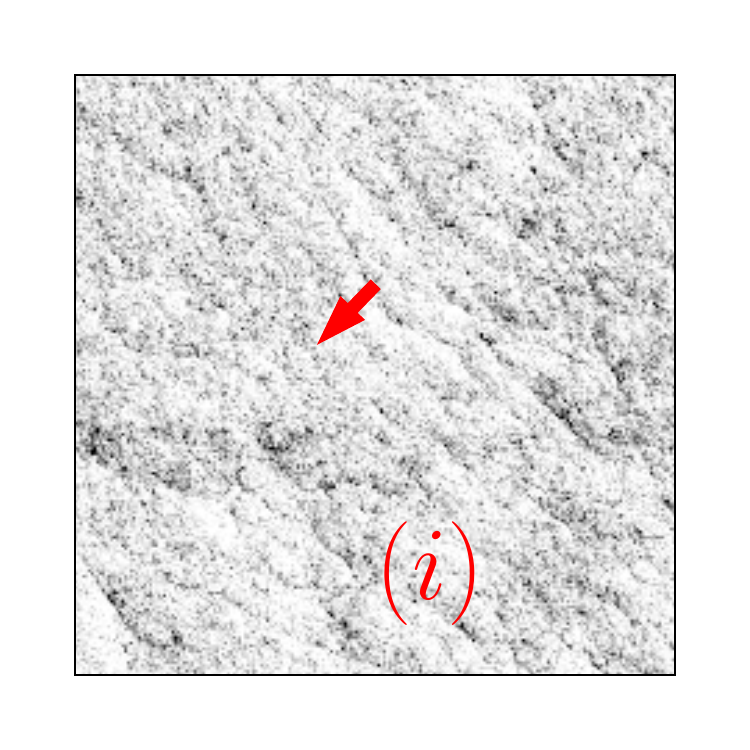}
	\includegraphics[width=0.22\textwidth]{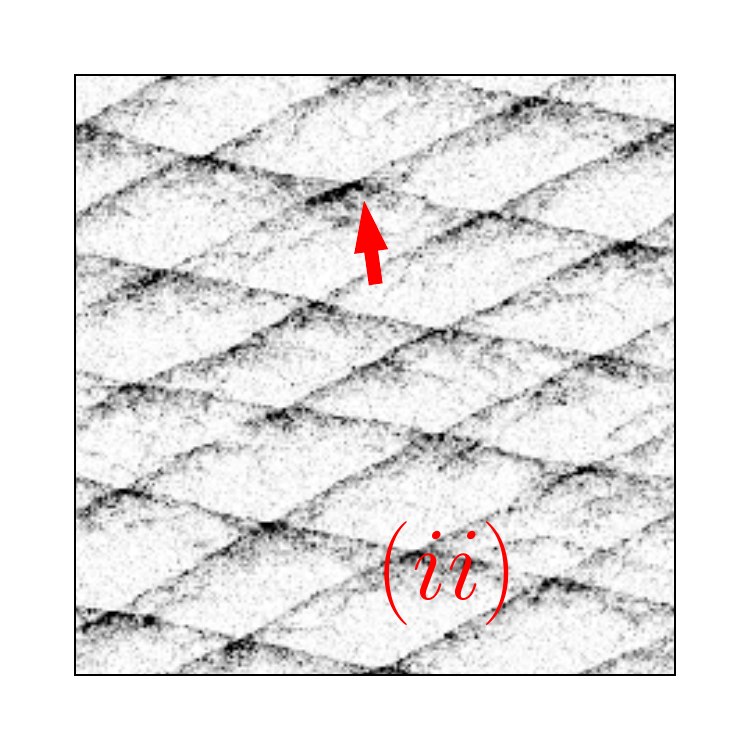}\\
	\includegraphics[width=0.22\textwidth]{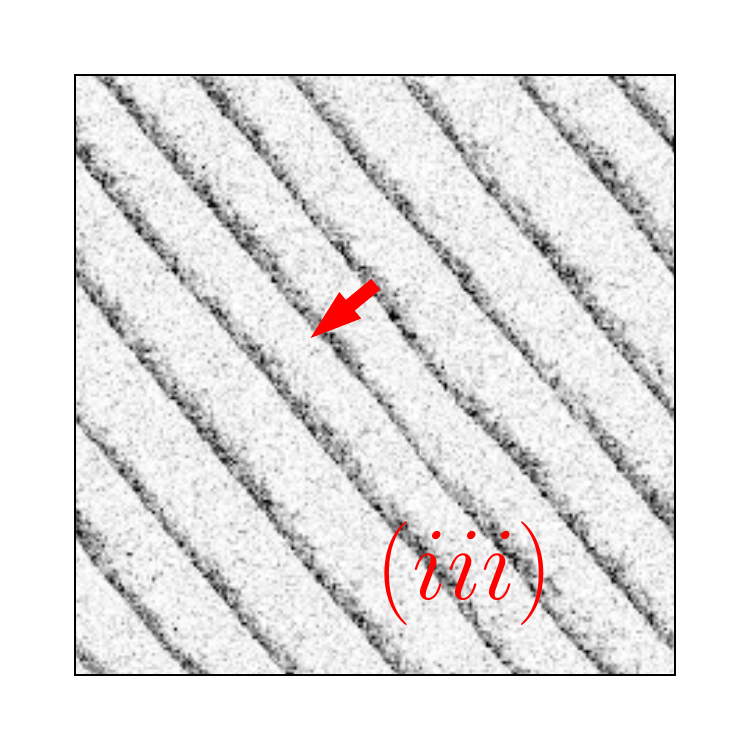}
	\includegraphics[width=0.22\textwidth]{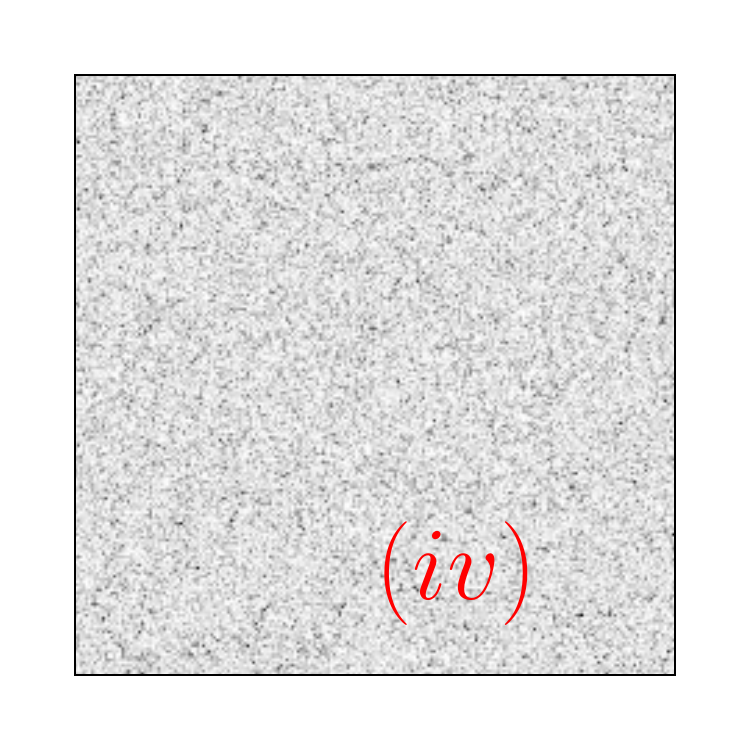}
	\caption{\label{fig:phases} Snapshots of the two dimensional standard Vicsek model in its four phases: $(i)$ Toner-Tu phase also called polar ordered liquid. $(ii)$ Cross sea phase. $(iii)$ Band phase. $(iv)$ Disordered phase. 
	Dark color represents high particle density.
	Red arrows indicate the average direction of motion.
	The systems have been started at random initial conditions. The snapshots have been taken after a thermalization time of $T=2\times10^5$. System parameters: $R=1$, $v=1$, $\Delta t=1$, particle number $N=10^6$, system size $L=1253.3$. Each particle has on average $2$ interacting neighbors. Periodic boundary conditions have been used.}
\end{figure}

However, there is also a large class of active systems where the fluid can be neglected, e.g. particles moving close to a surface which can transfer arbitrary amounts of momentum to the surrounding, and thus, momentum conservation is effectively not an issue.
Such systems with negligible fluid are called dry \cite{Chate20} and can be modeled by stochastic equations including positive and negative (activity) dissipation \cite{RBELS12}.
An important limiting case of strong activation and dissipation leads to a constant particle speed, an ingredient that is often directly incorporated in simplified models.
One of such models was introduced 25 years ago by Vicsek et. al. \cite{VCBCS95} and is one of the simplest and most studied models of active matter until today.
In the two-dimensional Vicsek model one considers particles that move with constant speed $v$ in individual directions given by angles $\theta_i$.
Those directions interact at discrete instances of time $n \Delta t$ and remain constant between those collisions. The interactions are given by the following rule
\begin{align}
	\theta_i(t + \Delta t) = \Theta\Bigg[\sum_{j\in \Omega_i(t+\Delta t)} \binom{\cos \theta_j(t)}{\sin \theta_j(t)}\Bigg]+ \xi_i(t),
	\label{eq:dynamics_vicsek}
\end{align}
where the set $\Omega_i$ contains the indexes of all particles $j$ that satisfy $|\mathbf{r}_i-\mathbf{r}_j|<R$ for some interaction radius $R$.
The function $\Theta(\mathbf{v})$ returns the angle that describes the direction of the two dimensional vector $\mathbf{v}$.
The $\xi_i(t)$ are independent random variables drawn uniformly from the interval $[-\pi\eta, \pi\eta]$. 
That is, after a discrete time interval $\Delta t$ all particles reorientate due to interactions. 
They take the average direction of all particles that are within distance $R$, disturbed by random noise of strength $\eta$. 
The model exhibits a transition towards collective motion for small noise (or large density) that was first believed to be continuous \cite{VCBCS95}. This shows the non-equilibrium nature of active matter, since in equilibrium such a transition would be strictly forbidden for short-range interactions in two dimensions \cite{MW66, Hohenberg66}.
It was found later, that for large enough systems the transition is actually discontinuous and goes along with the formation of high density bands that arrange regularly into waves \cite{GC04, CGGR08, Ihle13}. 
For even smaller noise strength (or higher density) there is another transition towards a homogeneous polar ordered phase that is also called Toner-Tu phase \cite{CGGR08, SCT15}. 
The behavior of the model has been described in analogy to a liquid-gas transition \cite{SCT15, SCBCT15}. The disordered phase at high noise intensities is considered as a gas, see Fig.~\ref{fig:phases} $(iv)$ for a snapshot of this phase. The phase of polar ordered bands is considered as  coexistence of a polar ordered liquid (the bands) and a disordered gas (the particles between the bands with almost no polar order), see Fig.~\ref{fig:phases} $(iii)$ and the Toner-Tu phase is considered as pure polar ordered liquid, see Fig.~\ref{fig:phases} $(i)$.
It was observed in phase $(iii)$ but close to phase $(i)$ that the bands do not achieve a smectic arrangement but they interact strongly and do not order, see Fig.~2b of Ref.~\cite{Chate20}.
It was explicitly formulated as a pending issue in \cite{Chate20} whether this system states should be considered separately from phase $(iii)$.

In this Letter, we answer this question demonstrating that the aforementioned parameter region represents another fourth phase of the Vicsek model, that has not been reported before to the best of the authors knowledge.
In Fig.~\ref{fig:phases} $(ii)$ we show a snapshot of this fourth phase which looks like a cross sea.
This is a phenomenon sometimes observed in the oceans when two wave systems like a swell (waves that are no longer under wind influence) and a wind sea (waves generated by wind) are combined, see e.g. \cite{OPT10}. 
It is considered to be particularly dangerous for ships, see e.g. \cite{TFBM05}.
Here however, this cross sea pattern is self-organized, since there is no external driving.
Furthermore, we demonstrate below that the cross sea pattern of the Vicsek model is not just a superposition of two planar waves.
We would also not expect this since nonlinear terms are important in the description of the Vicsek model.
In fact, we find that the particle density at the crossing points of the pattern is much higher than the sum of two bands.
Thus, particles accumulate at the crossing points of the pattern. Interestingly, similar effects seem to be present also in the cross sea in oceans, see \cite{OPT10} and references therein.  \phantom{\footnote{\label{foot:thermalization} For the smallest noise strength in the disordered phase, $\eta=0.45$, we still used $T=2\times 10^5$, for lager noise strengths, $\eta=0.46, \dots, 0.49$ we used a shorter time of $T=10^4$ since thermalization is much faster in the disordered phase.}}

We performed simulations for $N=10^6$ particles in a quadratic domain of size $L=1253.3$ with periodic boundary conditions.
Other system parameters are $R=\Delta t=v=1$ and the noise strength was varied from $\eta=0.2$ to $\eta=0.49$ in steps of $0.01$ simulating ten realizations for each noise strength. 
We made snapshots after $T=2\times 10^5$ thermalization time steps \footnotemark[1].
For the smallest noise strengths $\eta=0.20, \dots, 0.24$ we observe more or less homogeneous states. 
Starting from about $\eta=0.25$ structures are formed and for $\eta=0.27$ the first cross sea state arises. 
For $\eta=0.29, 0.30$ all observed realizations are clearly in a cross sea state.
Starting from $\eta=0.31$ some of the realizations are clearly cross sea and some others are clearly bands, whereas for $\eta=0.43, 0.44$ there are only band state realizations. 
Eventually, for $\eta \ge 0.45$ all realizations are disordered, see supplemental material at pages 6-31.
Hence, we observe three transitions between four different phases.
The observed mesh sizes and crossing angles of the cross sea pattern vary with parameters but also for different realizations of the same parameter set.
Possibly, they are affected by boundary conditions and the orientation of the average direction of motion with respect to the boundary might be important (for system sizes studied here).
To answer this question, further studies with significantly more data and a sophisticated image analysis are required.

A system configuration similar to the cross sea state has been shown very recently in Fig.~2b of \cite{Chate20}. 
There, it was described as strongly interacting bands that do not order.
In view of the results presented here, we can identify this state as very close to the transition between phases $(i)$ and $(ii)$.
It looks very similar to the states we find for $\eta=0.27$, see supplemental material.

To study the transitions in greater detail, we investigate a correlation order parameter that was recently introduced in \cite{KSZI20} and suggested to be used in the study of structural phase transitions, in particular out of equilibrium.
It is a local integral over the two particle correlation function formally given by
\begin{align}
	&C_2:= N^2 \int_{}^{} G_2(\mathbf{r}_1,\mathbf{r}_2)\diff \mathbf{r}_1 \, \diff \mathbf{r}_2 \, \theta(R-|\mathbf{r}_1|)\theta(R-|\mathbf{r}_2|),
	\label{eq:corrcircle_supp}
\end{align}
where $G_2(\mathbf{r}_1, \mathbf{r}_2):= P_2(\mathbf{r}_1, \mathbf{r}_2) - P_1(\mathbf{r}_1) P_1(\mathbf{r}_2)$ for one- and two-particle probability density functions $P_1$ and $P_2$, $\theta$ is the Heaviside function. 
For isotropic systems, the parameter can be expressed in terms of the usual pair correlation function $g(r)$ as
\begin{align}
	C_2= \bigg( \frac{N}{L^2}\bigg)^2 \int_{\mathbb{R}^2}^{}&[g(|\mathbf{r}_2-\mathbf{r}_1|)-1] 
	\label{eq:ctwo_paircorrelation}
	\\
	&\times \theta(R-|\mathbf{r}_1|) \theta(R-|\mathbf{r}_2|)\diff \mathbf{r}_1 \diff \mathbf{r}_2.
	\notag
\end{align}
Usually, this correlation parameter changes strongly when drastic spatial rearrangements occur.
Thus, it is appropriate to study the phase transition that we observe here.
Furthermore, it can be sampled efficiently, see \cite{KSZI20}.
\begin{figure}
	\includegraphics[width=0.45\textwidth]{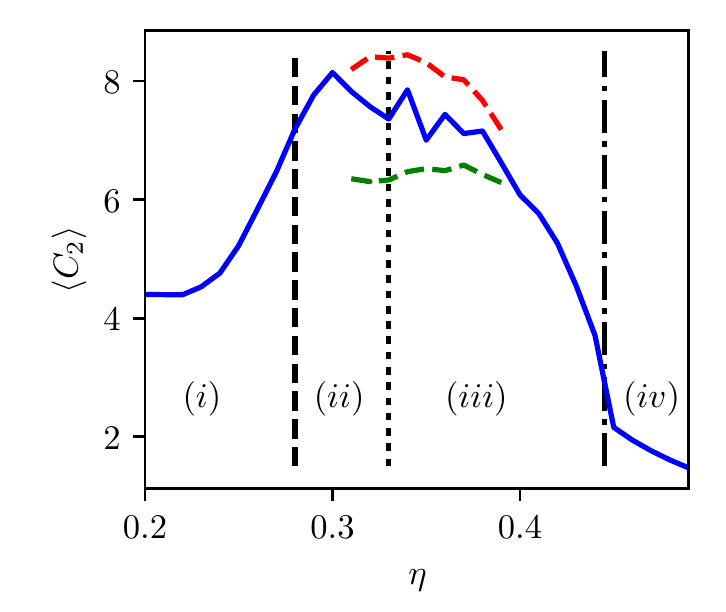}
	\caption{\label{fig:c2} Structural order parameter $C_2$ averaged over ten realizations and $10^4$ time steps after a thermalization time of $T=2\times 10^5$ [30] (blue solid line). For the noise values $\eta=0.31, \dots, 0.39$ we find several realizations in the cross sea phase as well as in the band phase, see supplemental material. Averaging only over realizations that are identified (by hand) as clearly in the cross sea phase results in the red upper dashed line. Analogously, averaging over the band phase realizations only results in the green lower dashed line. We see that the correlation parameter $C_2$ increases from the Toner-Tu phase $(i)$ to the cross sea phase $(ii)$ and than decreases to the band phase $(iii)$ and even more to the disordered phase $(iv)$. The transition lines from $(i)$ to $(ii)$ and $(ii)$ to $(iii)$ (vertical dashed and dotted lines) have been obtained as peaks of the Binder cumulant, see Fig.\ref{fig:binder}. The transition towards disorder (dash-dotted vertical line) was obtained by hand since all realizations are bands for $\eta=0.44$ and all realizations are disordered for $\eta=0.45$, see supplemental material. Parameters are as in Fig.~\ref{fig:phases}.}
\end{figure}

In Fig.~\ref{fig:c2} we show the average of the $C_2$ order parameter in dependence on the noise strength.
It increases drastically at the transition from phase $(i)$ to phase $(ii)$, then it decreases from phase $(ii)$ to phase $(iii)$ and decreases much more at the transition from phase $(iii)$ to phase $(iv)$.
In the average over all realizations (solid blue line) we cannot detect the transition between phases $(ii)$ and $(iii)$ that clearly, since for a relatively large noise range, we find realizations in both states, as discussed above.
However, if we measure the order parameter for realizations that show bands or cross sea states separately, we find significant differences in $C_2$, see dashed red and green lines in Fig.~\ref{fig:c2}.
The clear separation of the two dashed lines shows the discontinuous nature of the transition, which is also expected due to the different topological properties of the patterns.
\begin{figure}
	\includegraphics[width=0.45\textwidth]{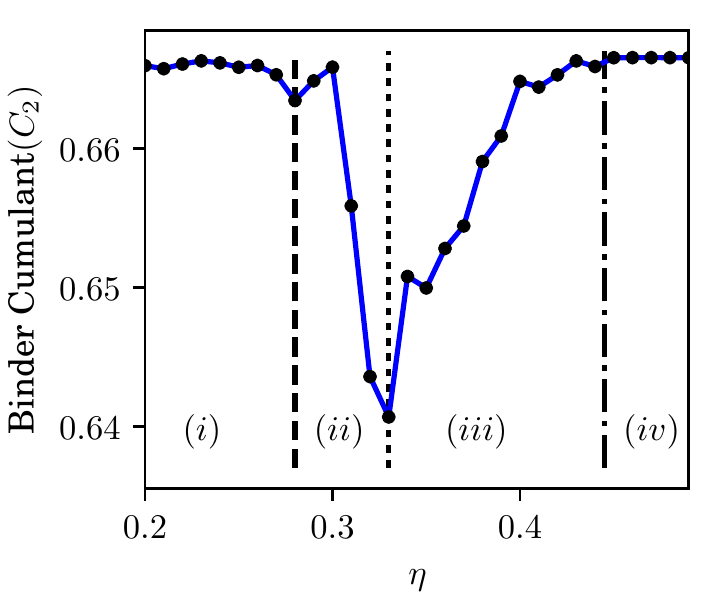}
	\caption{\label{fig:binder} Binder cumulant of the structural order parameter $C_2$ obtained from the same simulations as used in Fig.~\ref{fig:c2}.
	We can identify two clear peaks indicating the transitions form phases $(i)$ to $(ii)$ (dashed line) and from phases $(ii)$ to $(iii)$ (dotted line).
	In principle there should be another peak between phases $(iii)$ and $(iv)$ (dash-dotted line). However this is not observed here due to the rough resolution in $\eta$.}
\end{figure}

In order to justify whether the cross sea state is really a different phase, we measure the Binder cumulant of the $C_2$ order parameter.
In Fig.~\ref{fig:binder} we see that there are two peaks separating phases $(i)$ and $(ii)$, and $(ii)$ and $(iii)$, respectively.
In principle, we expect a third peak indicating the transition from phases $(iii)$ to $(iv)$.
However, it has been shown in \cite{KSZI20} that this peak is extremely sharp for large system sizes and thus not covered by the resolution of noise strength used here.
As this transition is not the major topic of this Letter, we just roughly estimated its position $\eta_{c3}\approx 0.445$ from looking at the snapshots and plotted it as the dash-dotted line in Fig.~\ref{fig:binder}.
The other transition noise strengths obtained from the peaks in the Binder cumulant, $\eta_{c1}=0.28$ and $\eta_{c2}=0.33$, agree well with the picture we got looking at the snapshots.
Remarkably, the peak between phases $(ii)$ and $(iii)$ is very broad. This can be easily understood, since we already observed that we find some realizations in both states over quite a large noise range.
Nevertheless, the peak clearly indicates a phase transition. We expect that the peak is narrowed significantly for much larger systems.
\begin{figure}
	\includegraphics[width=0.45\textwidth]{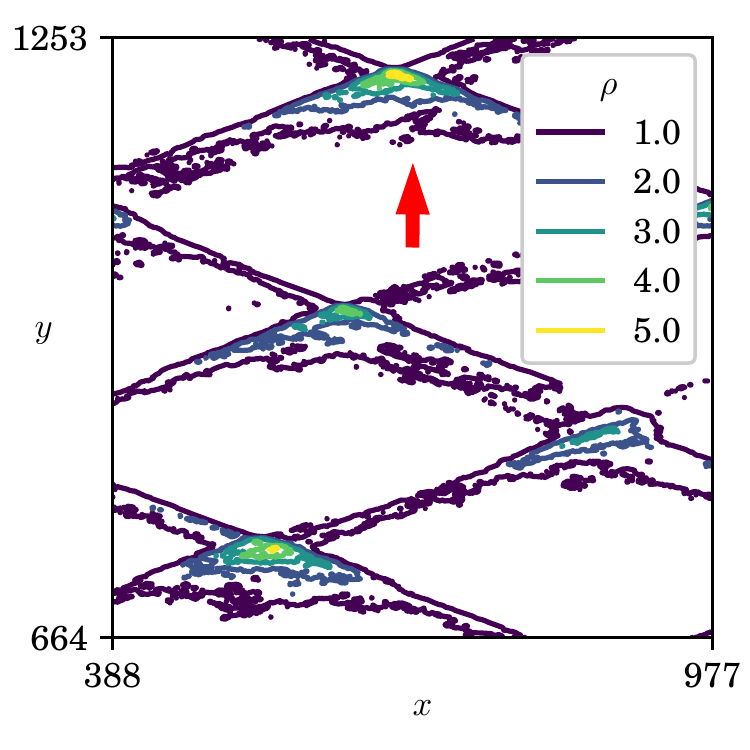}
	\caption{\label{fig:density} A section of the averaged particle density $\rho=N/L^2$ for one realization with $\eta=0.29$, displayed are five level lines.
	The averaging was performed over $10^3$ time steps in the co-moving frame of the pattern after a thermalization time of $T=2\times10^5$.
	The red arrow indicates the average direction of motion in the lab frame.
	We clearly see that a lot of mass accumulates at the crossing points of the pattern. 
	The crossing points densities are approximately $4-5$ times as large as the density along the single fronts. Parameters are as in Fig.~\ref{fig:phases}.}
\end{figure}

Looking on Fig.~\ref{fig:phases} $(ii)$ we might suppose that the cross sea state is just a superposition of two planar waves as they occur in phase $(iii)$.
To verify this hypothesis we measure the particle density averaged in the co-moving frame \footnote{To obtain the pattern co-moving frame we calculated at each time step a histogram of all particles positions before streaming. We then shifted the particle positions after streaming along the axis given by the mean velocity of all particles (among distances not greater than the particle speed $v$) and also made a histogram of the shifted after-streaming particle positions. We looked for the correlation between the histograms before streaming and after streaming with shift. The shift leading to the maximal correlation was then applied to all particles.} of the cross sea pattern.
One example is displayed in Fig.~\ref{fig:density}.
We observe that the density at the crossing points of the pattern is much larger than the sum of the densities of two fronts.
Hence, we conclude that the cross sea phase represents a standalone complex pattern and not just the sum of two waves.
\begin{figure}
	\includegraphics[width=0.45\textwidth]{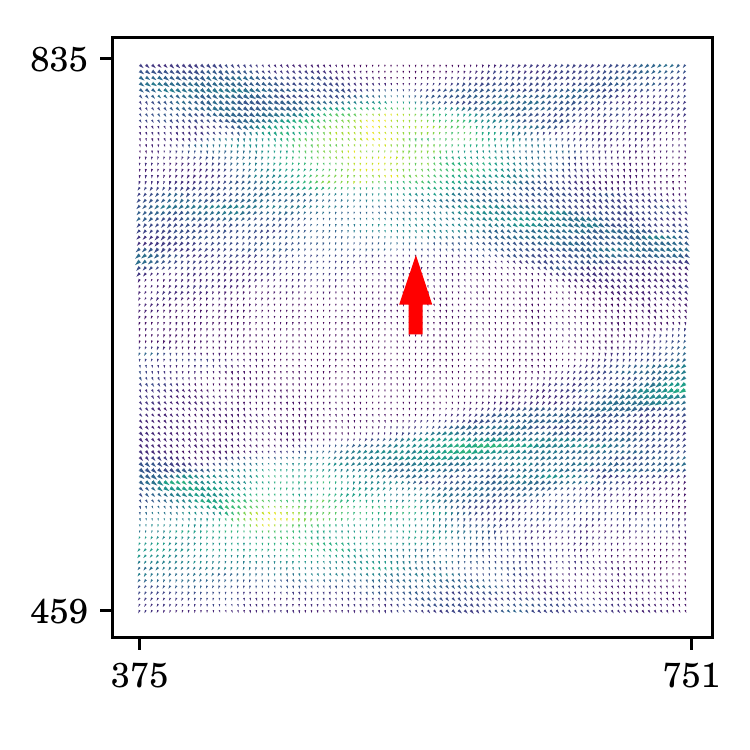}
	\caption{\label{fig:momentum} A section of the averaged momentum density for one realization with $\eta=0.29$.
	The maximum observed average momentum density (assuming unit mass of the particles) was $\approx 0.36$ at a density of $\rho\approx 1.32$ using a bin size of $\approx 4.2 \times 4.2$.
	The averaging was performed over $10^3$ time steps in the co-moving frame of the pattern after a thermalization time of $T=2\times10^5$.
	The red arrow indicates the average direction of motion in the lab frame.
	Colors encode the particle density (yellow=high, violet=low). 
	We observe that within the patterns co-moving frame mass is transported along the high density regions of the pattern on average in the direction opposite to the motion of the pattern in the lab frame.
	Parameters are as in Fig.~\ref{fig:phases}.}
\end{figure}

We also investigated the momentum density within the co-moving reference frame of the pattern.
We observe a non-zero particle flow along the band structures of the pattern in the direction opposite to the movement of the pattern in the lab frame, see Fig.~\ref{fig:momentum}.
The appearance of this mass flow is not too surprising since the pattern obeys no symmetry that would forbid such a flow.
Furthermore, in this non-equilibrium model we do not expect detailed balance to be present.

In summary, in this numerical study, we have shown that the two-dimensional standard Vicsek model forms complex self-organized cross sea patterns for very large system sizes and in certain parameter regimes.
We measured the density profile of the pattern and found that is not just a superposition of two waves but an independent structure.
We observe an interesting mass transport in the opposite direction compared to the pattern propagation.
Furthermore, measuring the Binder cumulant of a correlation order parameter, we have shown that the cross sea pattern represents a fourth phase of the Vicsek model.
There are two transitions from the cross sea phase: for lower noise intensity the system enters the Toner-Tu phase and for higher noise intensity it enters the phase of high density waves.
We thus answer a recently formulated question \cite{Chate20}.
On the other hand, an theoretical understanding of this novel phase is still missing.
Natural candidates for mathematical descriptions are field- or kinetic theories.
Both approaches seem to be challenging, since a full two-dimensional treatment is likely to be necessary in contrast to the band phase.
The appearance of the cross sea phase is likely to be relevant also for other models, similar to the Vicsek model. 
However, further studies are required.
Another pending question is whether an analogous phase exists in the three-dimensional Vicsek model.
A remarkable result from the general view on active matter is that apparently single species active systems can form complex patterns similar to those known from reaction-diffusion systems.

\begin{acknowledgments}
The authors gratefully acknowledges the GWK support for funding this project by providing computing time through the Center for Information Services and HPC (ZIH) at TU Dresden on the HRSK-II. The authors gratefully acknowledges the Universit\"atsrechenzentrum Greifswald for providing computing time.
\end{acknowledgments}

\newpage
\onecolumngrid
\newpage
\appendix
\section{Supplemental material: Snapshots of ten realizations, all started from random initial conditions, parameters as in the letter.}
$\eta=0.20$.
\begin{figure}[H]
\begin{center}
 \includegraphics[width=0.28\textwidth]{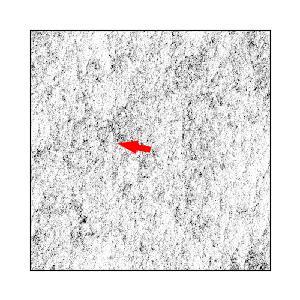}
 \includegraphics[width=0.28\textwidth]{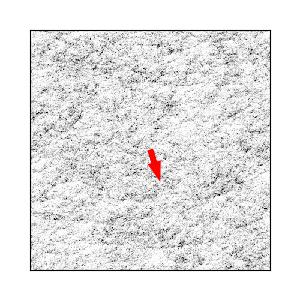}
 \includegraphics[width=0.28\textwidth]{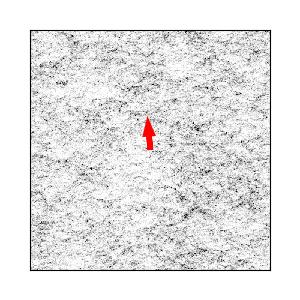}\\
 \includegraphics[width=0.28\textwidth]{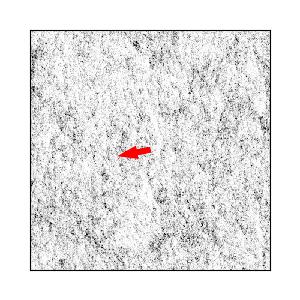}
 \includegraphics[width=0.28\textwidth]{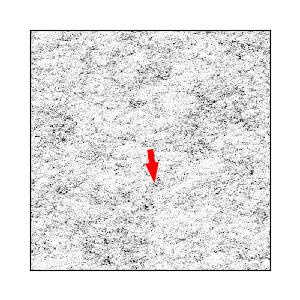}
 \includegraphics[width=0.28\textwidth]{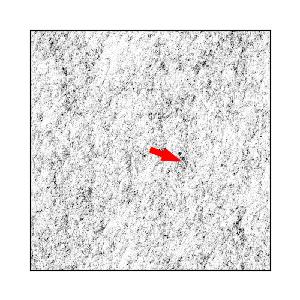}\\
 \includegraphics[width=0.28\textwidth]{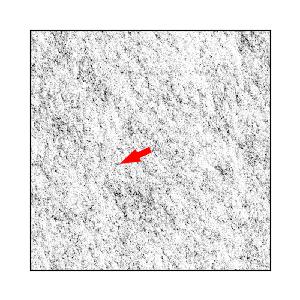}
 \includegraphics[width=0.28\textwidth]{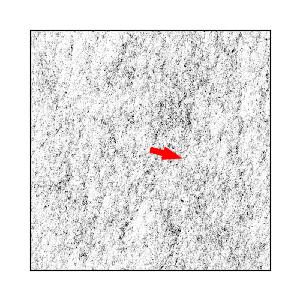}
 \includegraphics[width=0.28\textwidth]{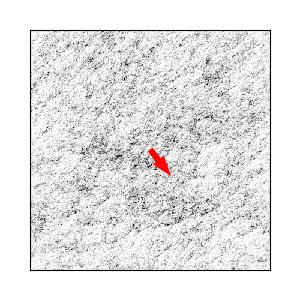}\\
\includegraphics[width=0.28\textwidth]{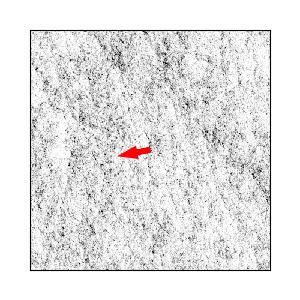}
\end{center}
\end{figure}

\newpage
$\eta=0.21$.
\begin{figure}[H]
\begin{center}
 \includegraphics[width=0.28\textwidth]{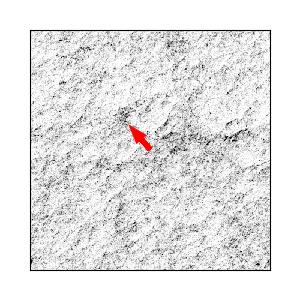}
 \includegraphics[width=0.28\textwidth]{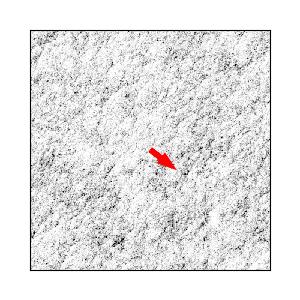}
 \includegraphics[width=0.28\textwidth]{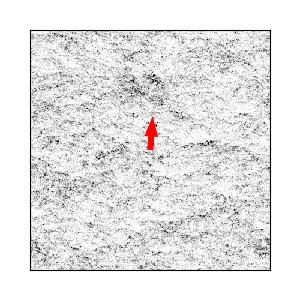}\\
 \includegraphics[width=0.28\textwidth]{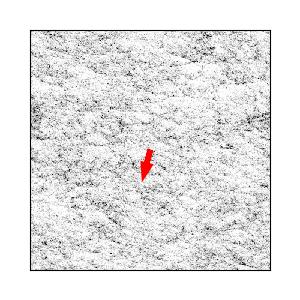}
 \includegraphics[width=0.28\textwidth]{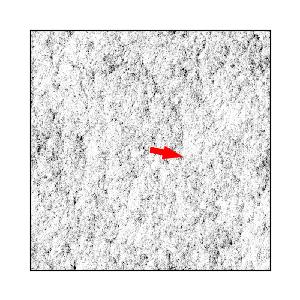}
 \includegraphics[width=0.28\textwidth]{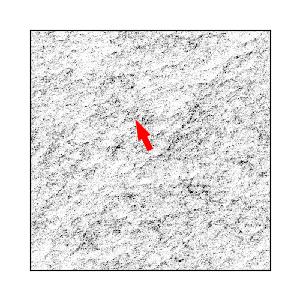}\\
 \includegraphics[width=0.28\textwidth]{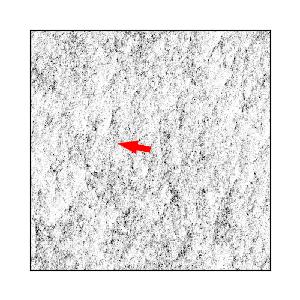}
 \includegraphics[width=0.28\textwidth]{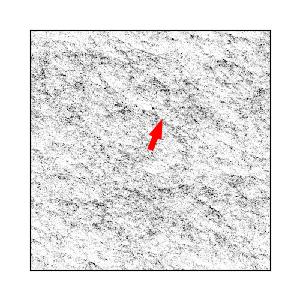}
 \includegraphics[width=0.28\textwidth]{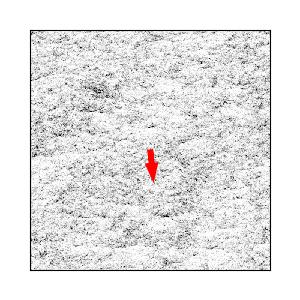}\\
\includegraphics[width=0.28\textwidth]{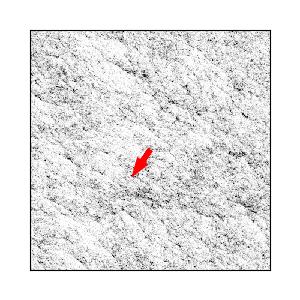}
\end{center}
\end{figure}

\newpage
$\eta=0.22$.
\begin{figure}[H]
\begin{center}
 \includegraphics[width=0.28\textwidth]{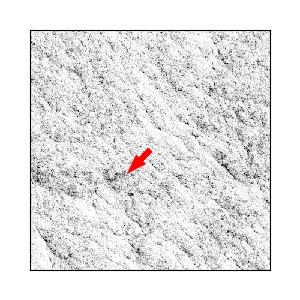}
 \includegraphics[width=0.28\textwidth]{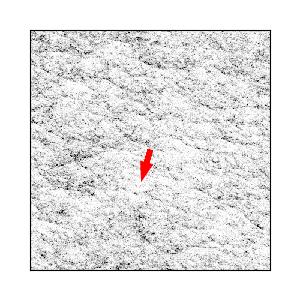}
 \includegraphics[width=0.28\textwidth]{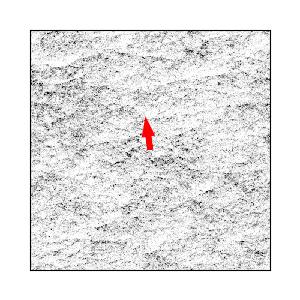}\\
 \includegraphics[width=0.28\textwidth]{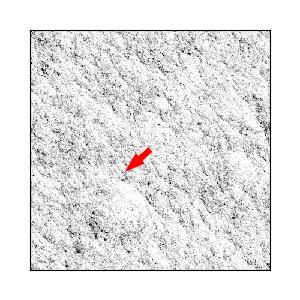}
 \includegraphics[width=0.28\textwidth]{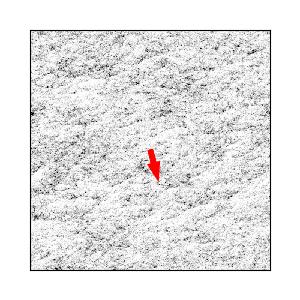}
 \includegraphics[width=0.28\textwidth]{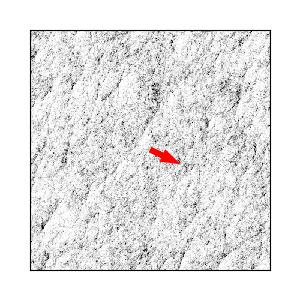}\\
 \includegraphics[width=0.28\textwidth]{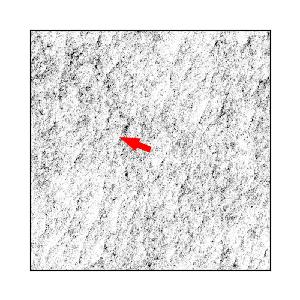}
 \includegraphics[width=0.28\textwidth]{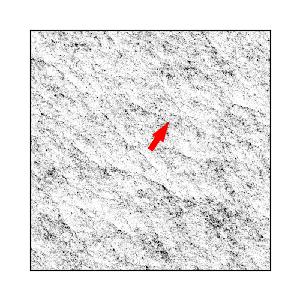}
 \includegraphics[width=0.28\textwidth]{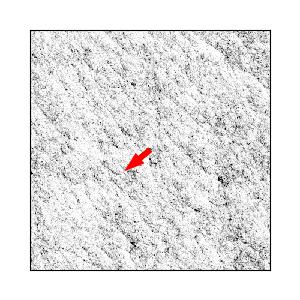}\\
\includegraphics[width=0.28\textwidth]{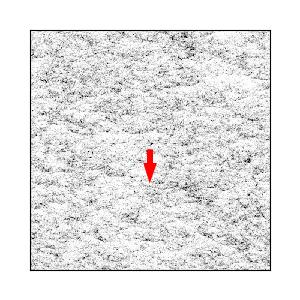}
\end{center}
\end{figure}

\newpage
$\eta=0.23$.
\begin{figure}[H]
\begin{center}
 \includegraphics[width=0.28\textwidth]{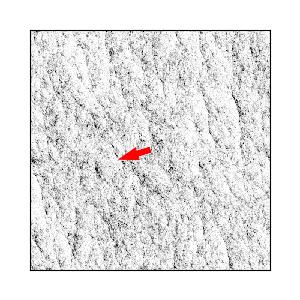}
 \includegraphics[width=0.28\textwidth]{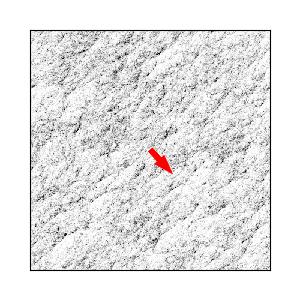}
 \includegraphics[width=0.28\textwidth]{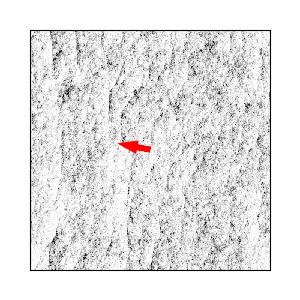}\\
 \includegraphics[width=0.28\textwidth]{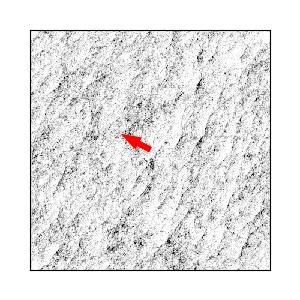}
 \includegraphics[width=0.28\textwidth]{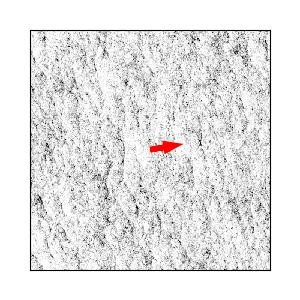}
 \includegraphics[width=0.28\textwidth]{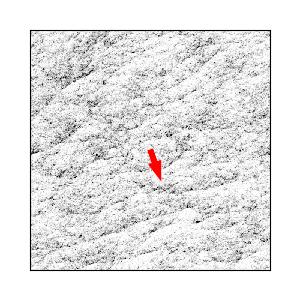}\\
 \includegraphics[width=0.28\textwidth]{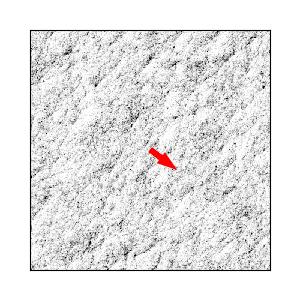}
 \includegraphics[width=0.28\textwidth]{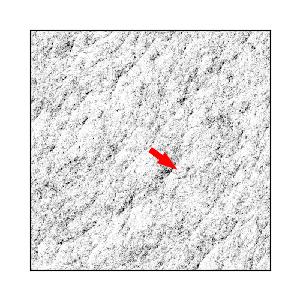}
 \includegraphics[width=0.28\textwidth]{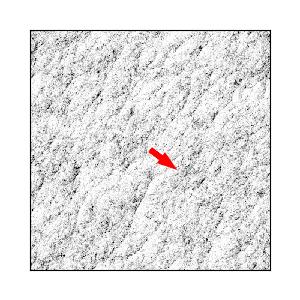}\\
\includegraphics[width=0.28\textwidth]{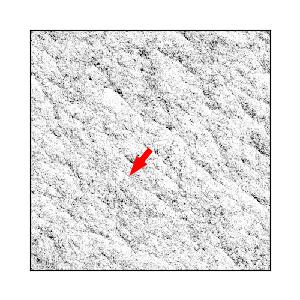}
\end{center}
\end{figure}

\newpage
$\eta=0.24$.
\begin{figure}[H]
\begin{center}
 \includegraphics[width=0.28\textwidth]{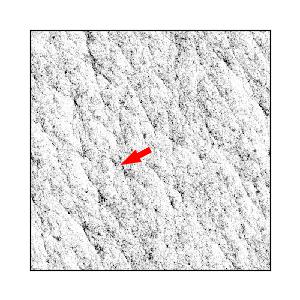}
 \includegraphics[width=0.28\textwidth]{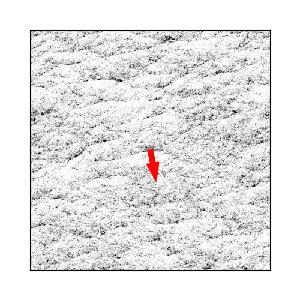}
 \includegraphics[width=0.28\textwidth]{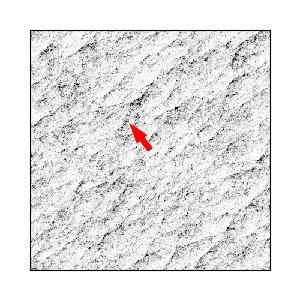}\\
 \includegraphics[width=0.28\textwidth]{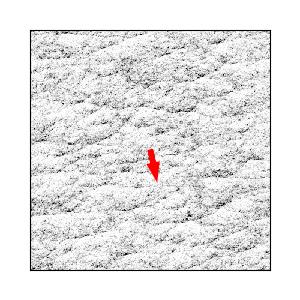}
 \includegraphics[width=0.28\textwidth]{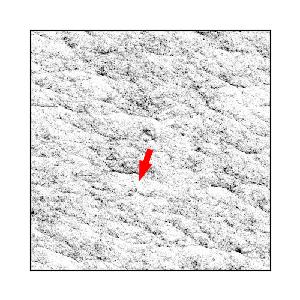}
 \includegraphics[width=0.28\textwidth]{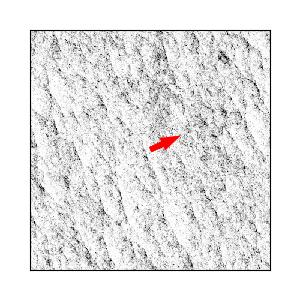}\\
 \includegraphics[width=0.28\textwidth]{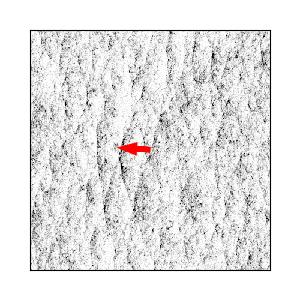}
 \includegraphics[width=0.28\textwidth]{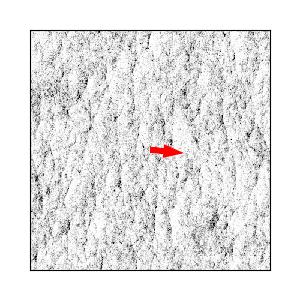}
 \includegraphics[width=0.28\textwidth]{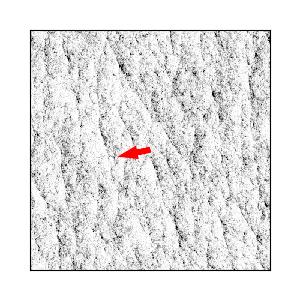}\\
\includegraphics[width=0.28\textwidth]{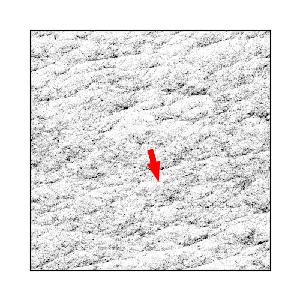}
\end{center}
\end{figure}

\newpage
$\eta=0.25$.
\begin{figure}[H]
\begin{center}
 \includegraphics[width=0.28\textwidth]{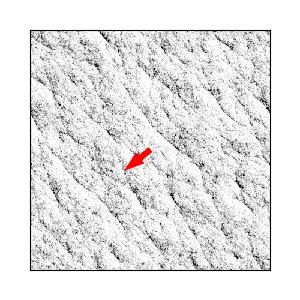}
 \includegraphics[width=0.28\textwidth]{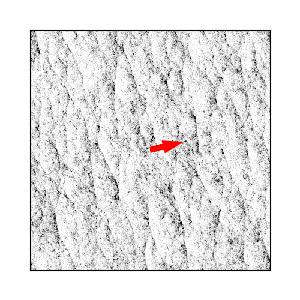}
 \includegraphics[width=0.28\textwidth]{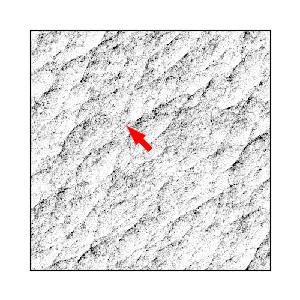}\\
 \includegraphics[width=0.28\textwidth]{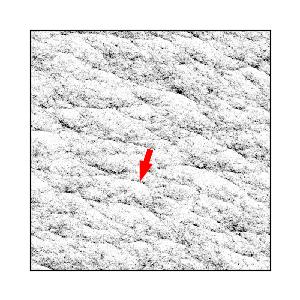}
 \includegraphics[width=0.28\textwidth]{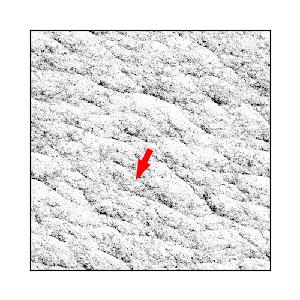}
 \includegraphics[width=0.28\textwidth]{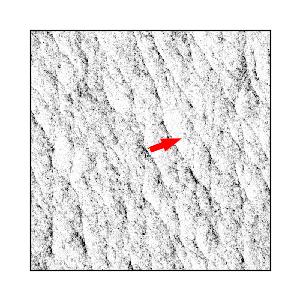}\\
 \includegraphics[width=0.28\textwidth]{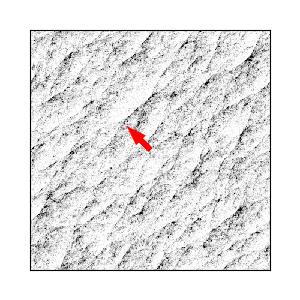}
 \includegraphics[width=0.28\textwidth]{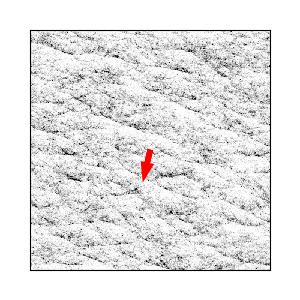}
 \includegraphics[width=0.28\textwidth]{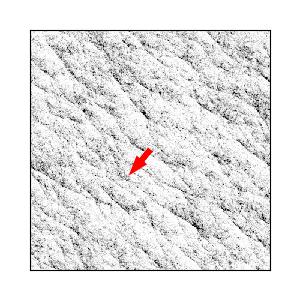}\\
\includegraphics[width=0.28\textwidth]{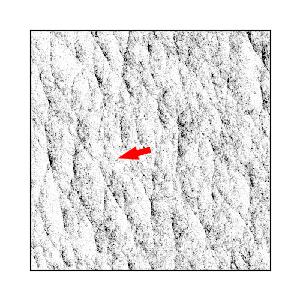}
\end{center}
\end{figure}

\newpage
$\eta=0.26$.
\begin{figure}[H]
\begin{center}
 \includegraphics[width=0.28\textwidth]{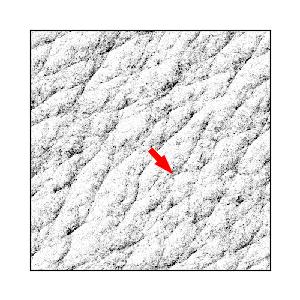}
 \includegraphics[width=0.28\textwidth]{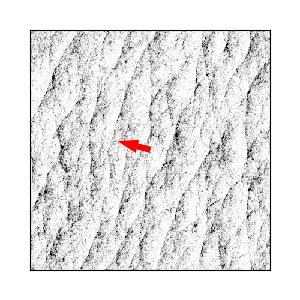}
 \includegraphics[width=0.28\textwidth]{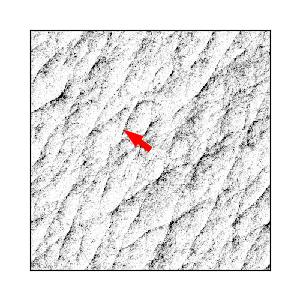}\\
 \includegraphics[width=0.28\textwidth]{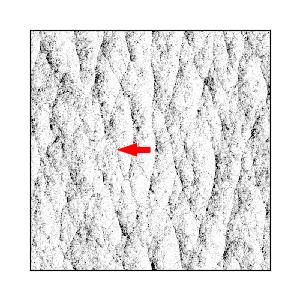}
 \includegraphics[width=0.28\textwidth]{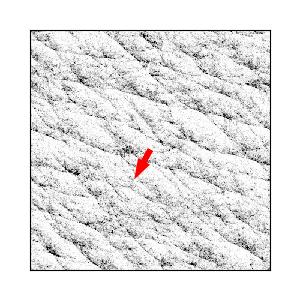}
 \includegraphics[width=0.28\textwidth]{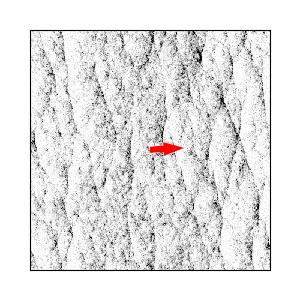}\\
 \includegraphics[width=0.28\textwidth]{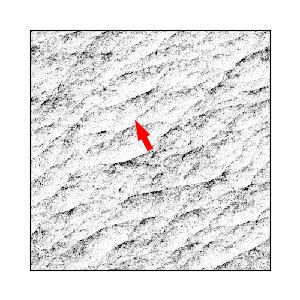}
 \includegraphics[width=0.28\textwidth]{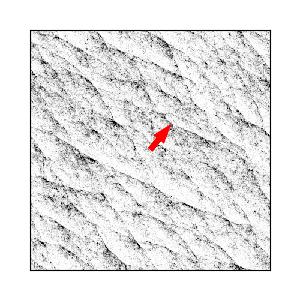}
 \includegraphics[width=0.28\textwidth]{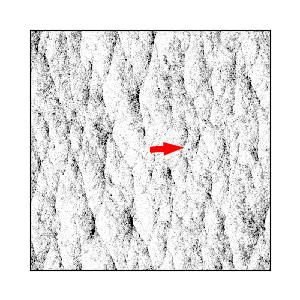}\\
\includegraphics[width=0.28\textwidth]{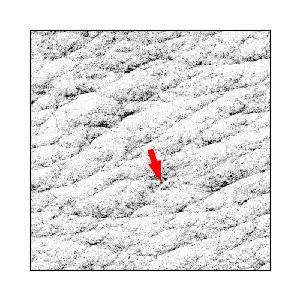}
\end{center}
\end{figure}

\newpage
$\eta=0.27$.
\begin{figure}[H]
\begin{center}
 \includegraphics[width=0.28\textwidth]{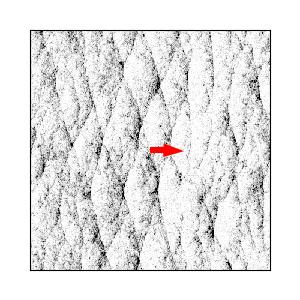}
 \includegraphics[width=0.28\textwidth]{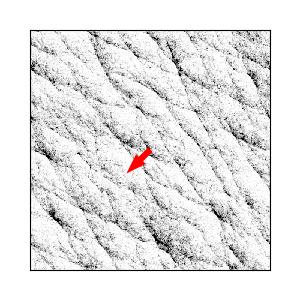}
 \includegraphics[width=0.28\textwidth]{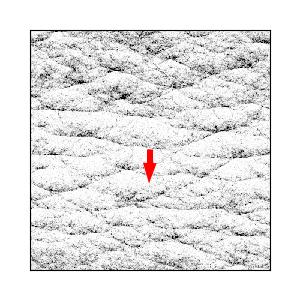}\\
 \includegraphics[width=0.28\textwidth]{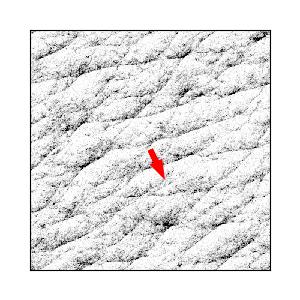}
 \includegraphics[width=0.28\textwidth]{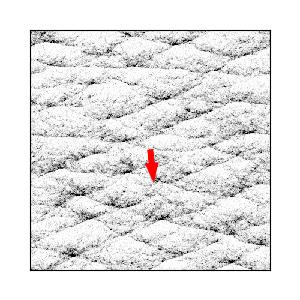}
 \includegraphics[width=0.28\textwidth]{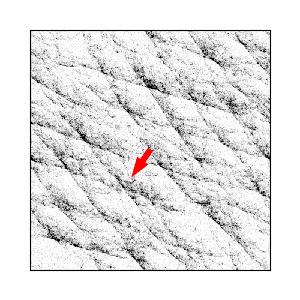}\\
 \includegraphics[width=0.28\textwidth]{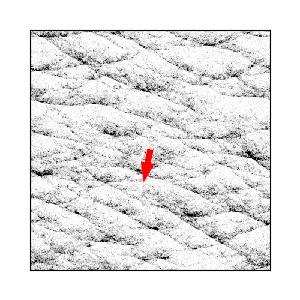}
 \includegraphics[width=0.28\textwidth]{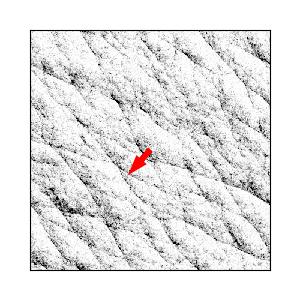}
 \includegraphics[width=0.28\textwidth]{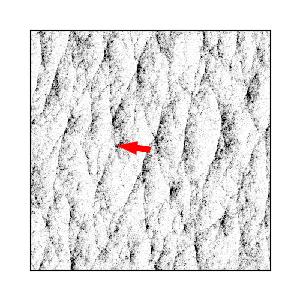}\\
\includegraphics[width=0.28\textwidth]{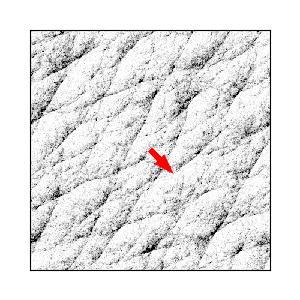}
\end{center}
\end{figure}

\newpage
$\eta=0.28$.
\begin{figure}[H]
\begin{center}
 \includegraphics[width=0.28\textwidth]{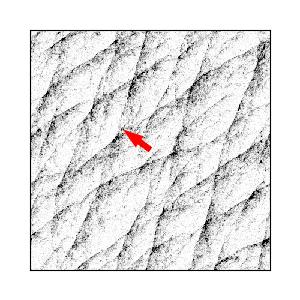}
 \includegraphics[width=0.28\textwidth]{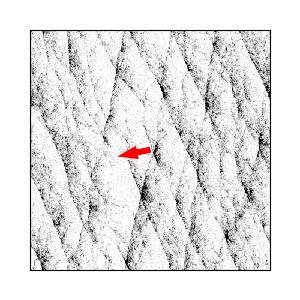}
 \includegraphics[width=0.28\textwidth]{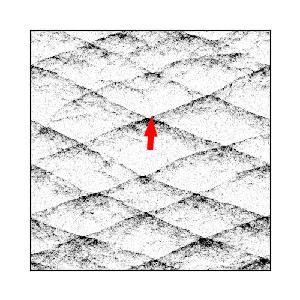}\\
 \includegraphics[width=0.28\textwidth]{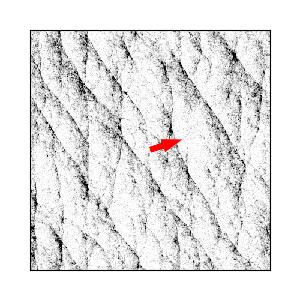}
 \includegraphics[width=0.28\textwidth]{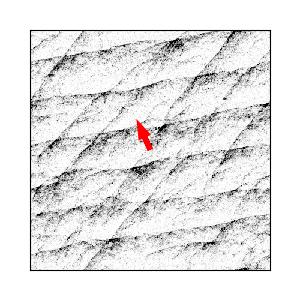}
 \includegraphics[width=0.28\textwidth]{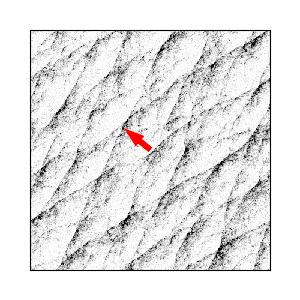}\\
 \includegraphics[width=0.28\textwidth]{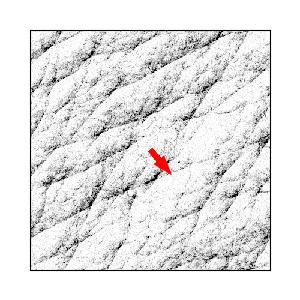}
 \includegraphics[width=0.28\textwidth]{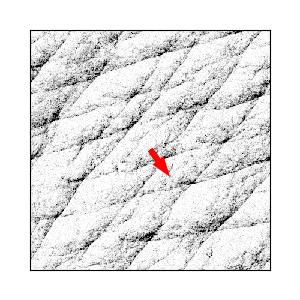}
 \includegraphics[width=0.28\textwidth]{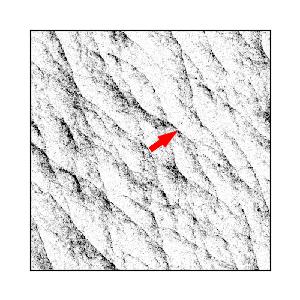}\\
\includegraphics[width=0.28\textwidth]{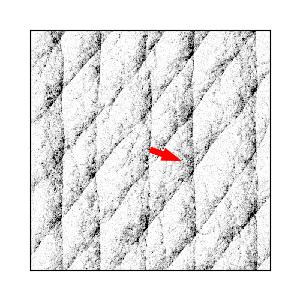}
\end{center}
\end{figure}

\newpage
$\eta=0.29$.
\begin{figure}[H]
\begin{center}
 \includegraphics[width=0.28\textwidth]{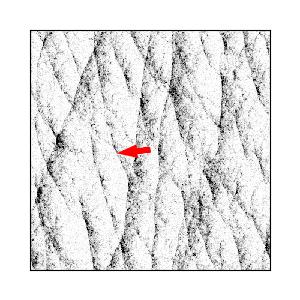}
 \includegraphics[width=0.28\textwidth]{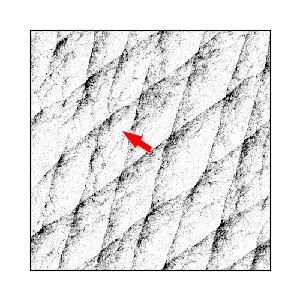}
 \includegraphics[width=0.28\textwidth]{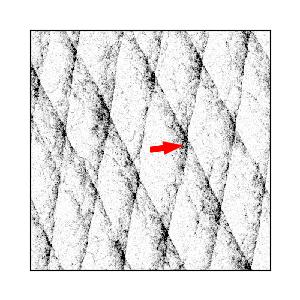}\\
 \includegraphics[width=0.28\textwidth]{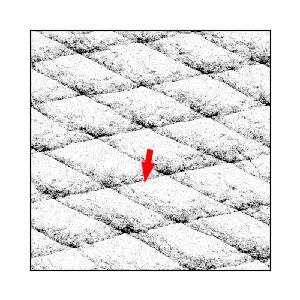}
 \includegraphics[width=0.28\textwidth]{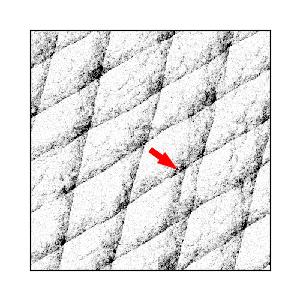}
 \includegraphics[width=0.28\textwidth]{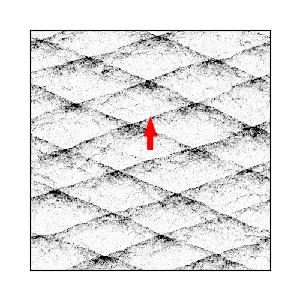}\\
 \includegraphics[width=0.28\textwidth]{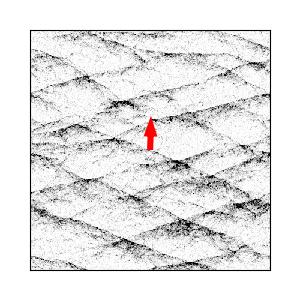}
 \includegraphics[width=0.28\textwidth]{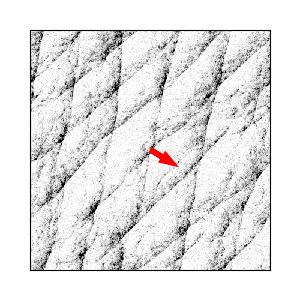}
 \includegraphics[width=0.28\textwidth]{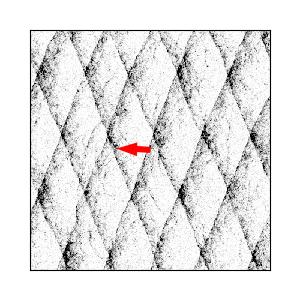}\\
\includegraphics[width=0.28\textwidth]{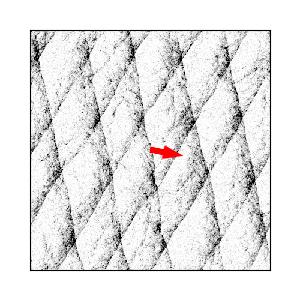}
\end{center}
\end{figure}

\newpage
$\eta=0.30$.
\begin{figure}[H]
\begin{center}
 \includegraphics[width=0.28\textwidth]{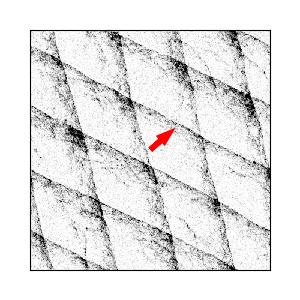}
 \includegraphics[width=0.28\textwidth]{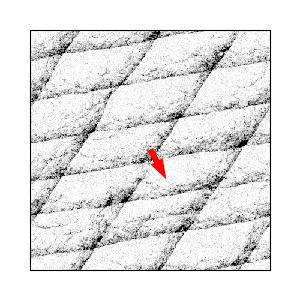}
 \includegraphics[width=0.28\textwidth]{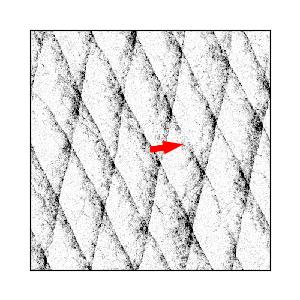}\\
 \includegraphics[width=0.28\textwidth]{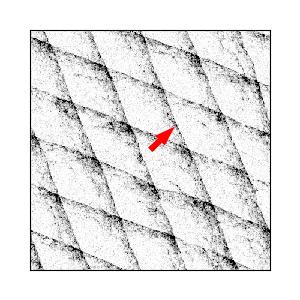}
 \includegraphics[width=0.28\textwidth]{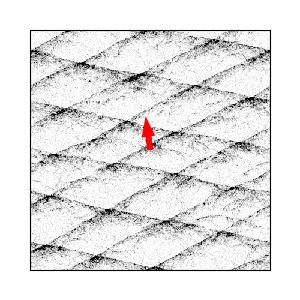}
 \includegraphics[width=0.28\textwidth]{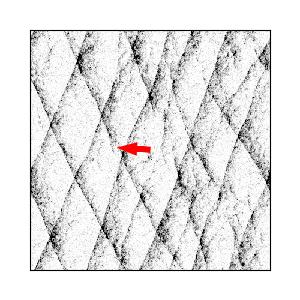}\\
 \includegraphics[width=0.28\textwidth]{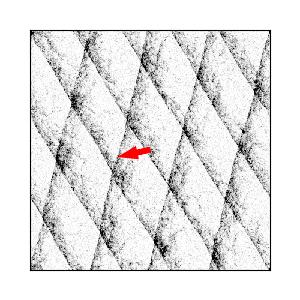}
 \includegraphics[width=0.28\textwidth]{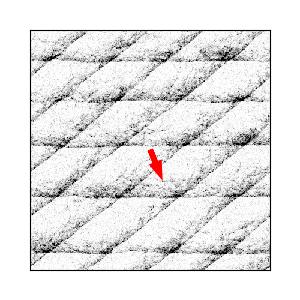}
 \includegraphics[width=0.28\textwidth]{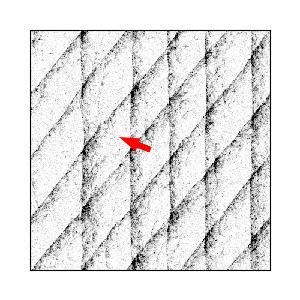}\\
\includegraphics[width=0.28\textwidth]{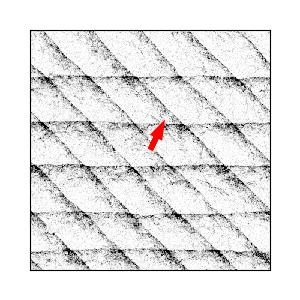}
\end{center}
\end{figure}

\newpage
$\eta=0.31$.
\begin{figure}[H]
\begin{center}
 \includegraphics[width=0.28\textwidth]{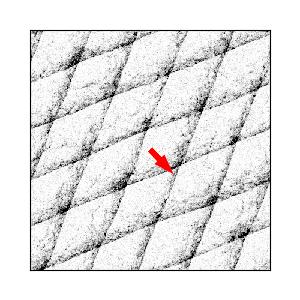}
 \includegraphics[width=0.28\textwidth]{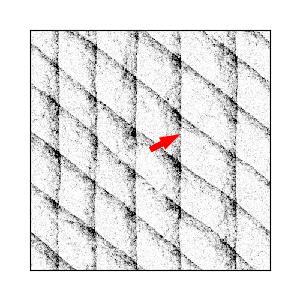}
 \includegraphics[width=0.28\textwidth]{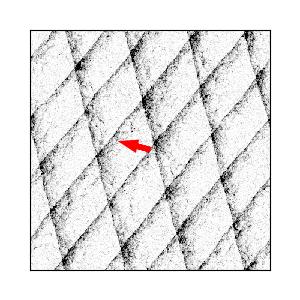}\\
 \includegraphics[width=0.28\textwidth]{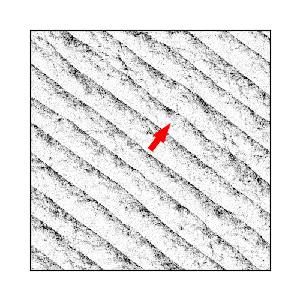}
 \includegraphics[width=0.28\textwidth]{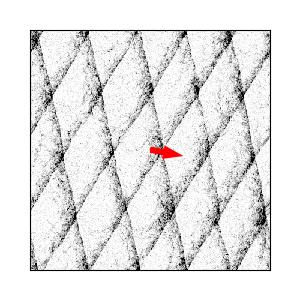}
 \includegraphics[width=0.28\textwidth]{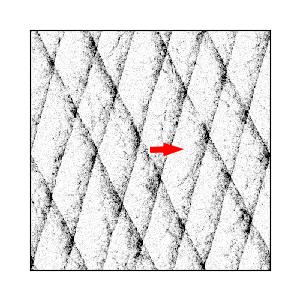}\\
 \includegraphics[width=0.28\textwidth]{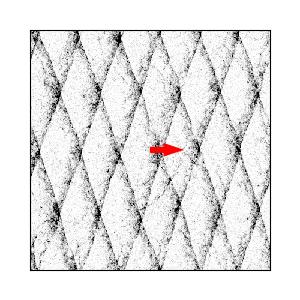}
 \includegraphics[width=0.28\textwidth]{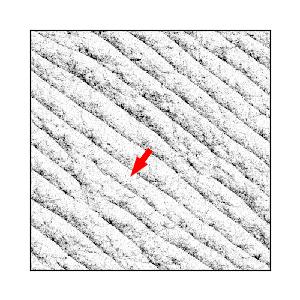}
 \includegraphics[width=0.28\textwidth]{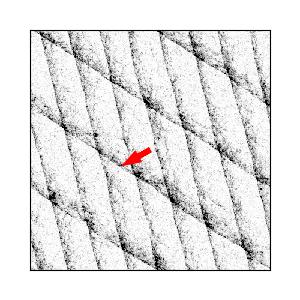}\\
\includegraphics[width=0.28\textwidth]{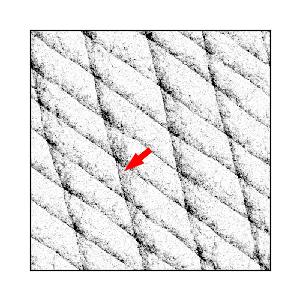}
\end{center}
\end{figure}

\newpage
$\eta=0.32$.
\begin{figure}[H]
\begin{center}
 \includegraphics[width=0.28\textwidth]{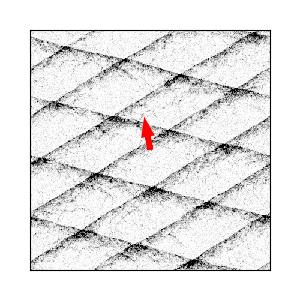}
 \includegraphics[width=0.28\textwidth]{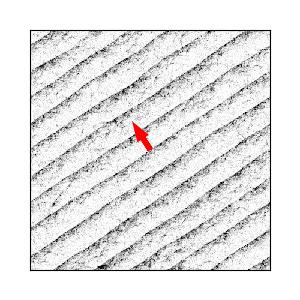}
 \includegraphics[width=0.28\textwidth]{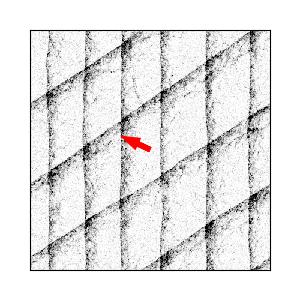}\\
 \includegraphics[width=0.28\textwidth]{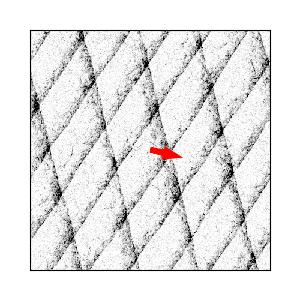}
 \includegraphics[width=0.28\textwidth]{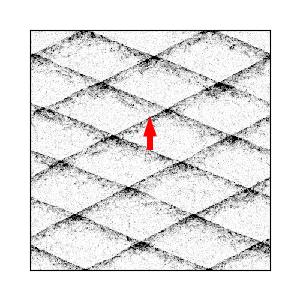}
 \includegraphics[width=0.28\textwidth]{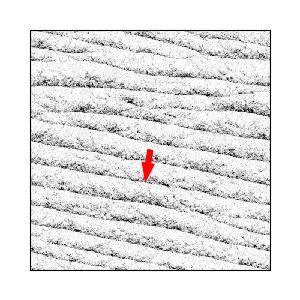}\\
 \includegraphics[width=0.28\textwidth]{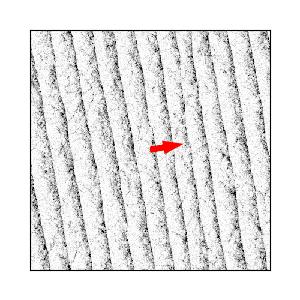}
 \includegraphics[width=0.28\textwidth]{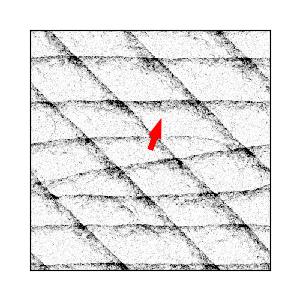}
 \includegraphics[width=0.28\textwidth]{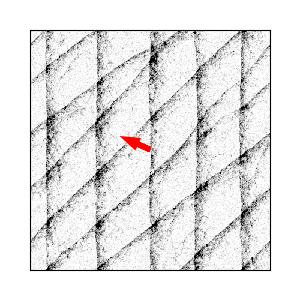}\\
\includegraphics[width=0.28\textwidth]{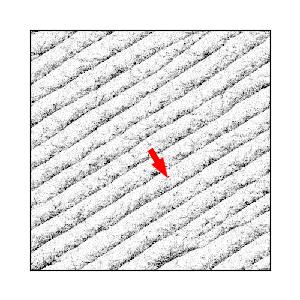}
\end{center}
\end{figure}

\newpage
$\eta=0.33$.
\begin{figure}[H]
\begin{center}
 \includegraphics[width=0.28\textwidth]{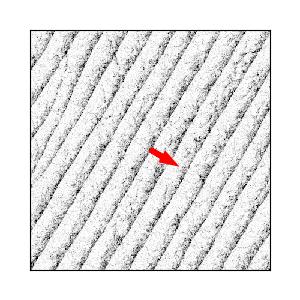}
 \includegraphics[width=0.28\textwidth]{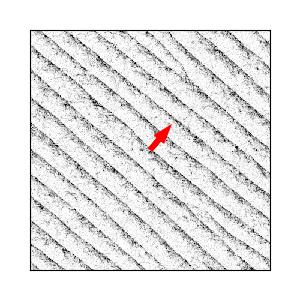}
 \includegraphics[width=0.28\textwidth]{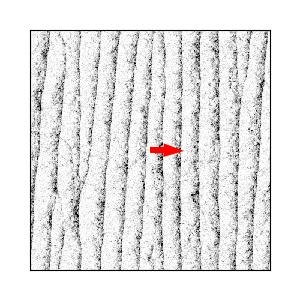}\\
 \includegraphics[width=0.28\textwidth]{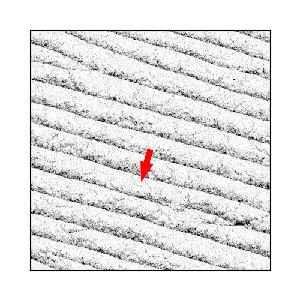}
 \includegraphics[width=0.28\textwidth]{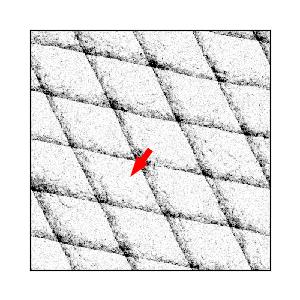}
 \includegraphics[width=0.28\textwidth]{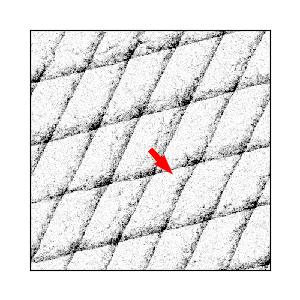}\\
 \includegraphics[width=0.28\textwidth]{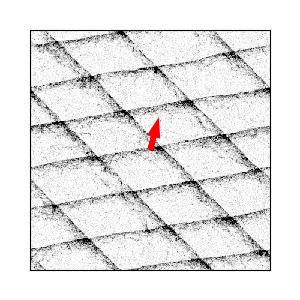}
 \includegraphics[width=0.28\textwidth]{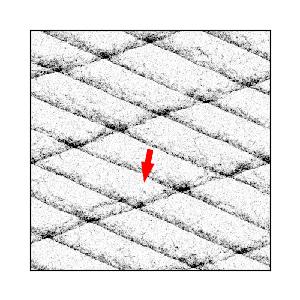}
 \includegraphics[width=0.28\textwidth]{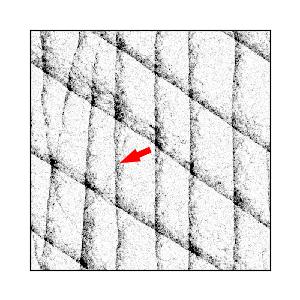}\\
\includegraphics[width=0.28\textwidth]{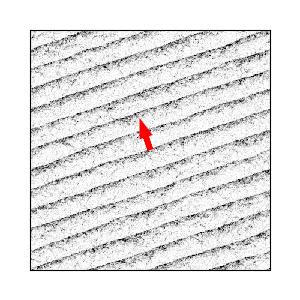}
\end{center}
\end{figure}

\newpage
$\eta=0.34$.
\begin{figure}[H]
\begin{center}
 \includegraphics[width=0.28\textwidth]{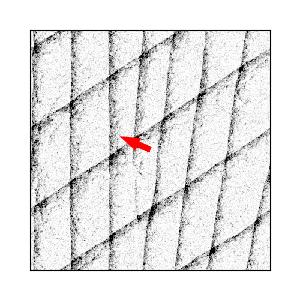}
 \includegraphics[width=0.28\textwidth]{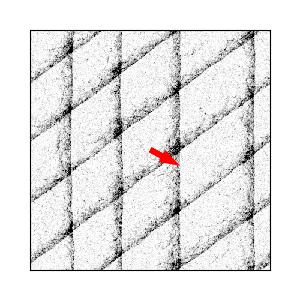}
 \includegraphics[width=0.28\textwidth]{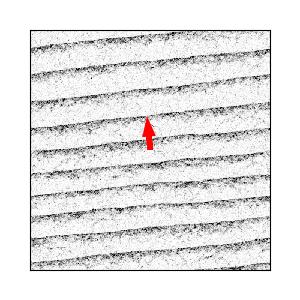}\\
 \includegraphics[width=0.28\textwidth]{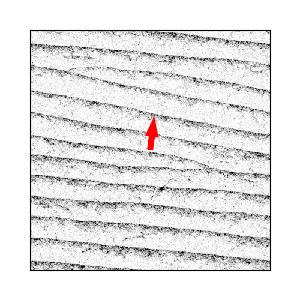}
 \includegraphics[width=0.28\textwidth]{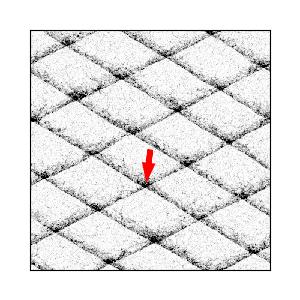}
 \includegraphics[width=0.28\textwidth]{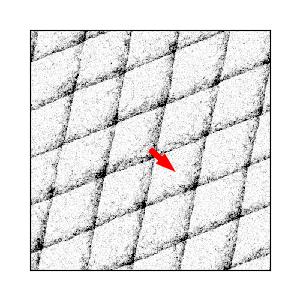}\\
 \includegraphics[width=0.28\textwidth]{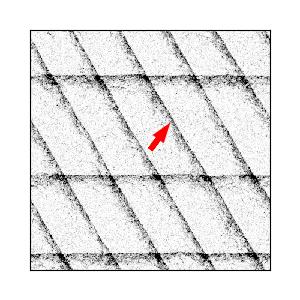}
 \includegraphics[width=0.28\textwidth]{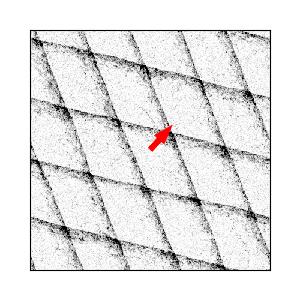}
 \includegraphics[width=0.28\textwidth]{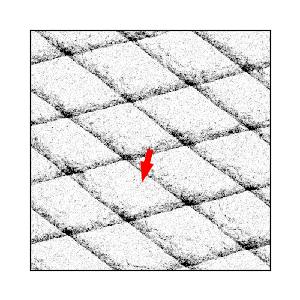}\\
\includegraphics[width=0.28\textwidth]{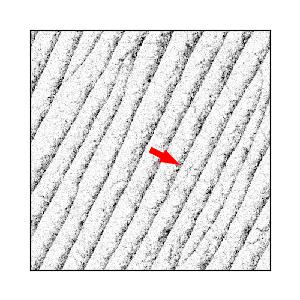}
\end{center}
\end{figure}

\newpage
$\eta=0.35$.
\begin{figure}[H]
\begin{center}
 \includegraphics[width=0.28\textwidth]{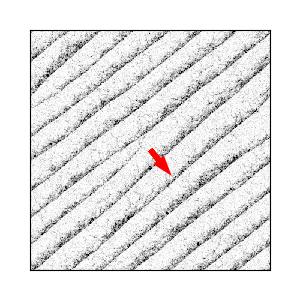}
 \includegraphics[width=0.28\textwidth]{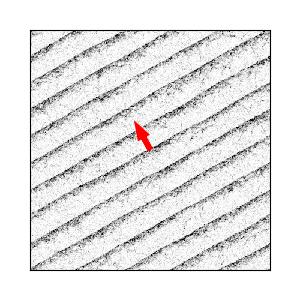}
 \includegraphics[width=0.28\textwidth]{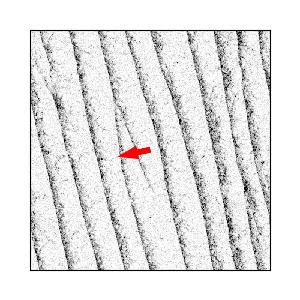}\\
 \includegraphics[width=0.28\textwidth]{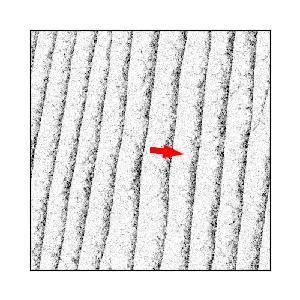}
 \includegraphics[width=0.28\textwidth]{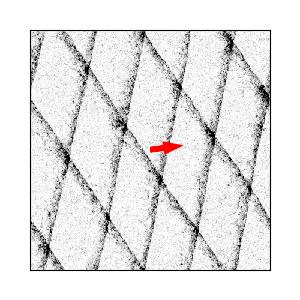}
 \includegraphics[width=0.28\textwidth]{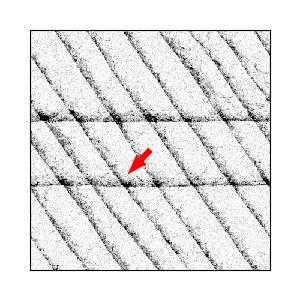}\\
 \includegraphics[width=0.28\textwidth]{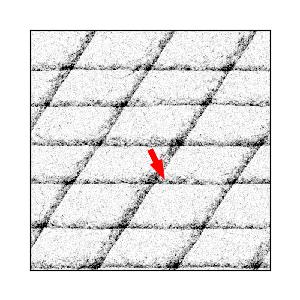}
 \includegraphics[width=0.28\textwidth]{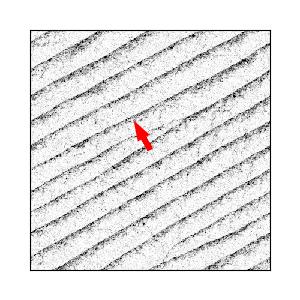}
 \includegraphics[width=0.28\textwidth]{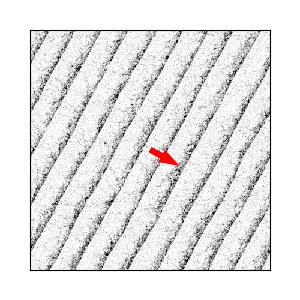}\\
\includegraphics[width=0.28\textwidth]{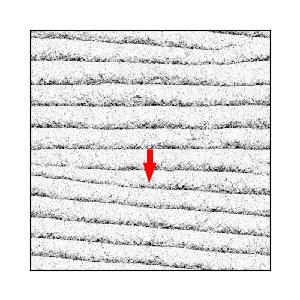}
\end{center}
\end{figure}

\newpage
$\eta=0.36$.
\begin{figure}[H]
\begin{center}
 \includegraphics[width=0.28\textwidth]{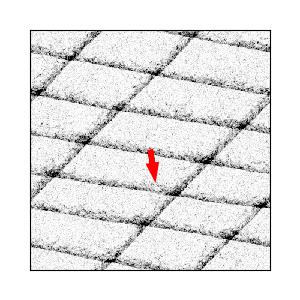}
 \includegraphics[width=0.28\textwidth]{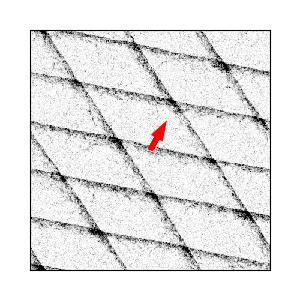}
 \includegraphics[width=0.28\textwidth]{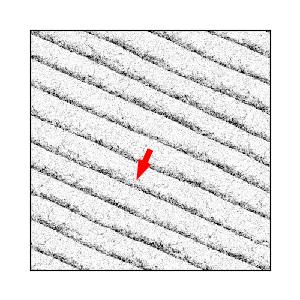}\\
 \includegraphics[width=0.28\textwidth]{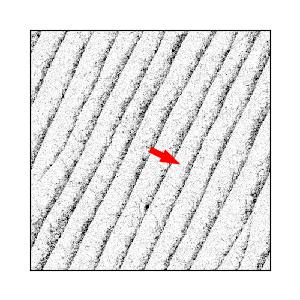}
 \includegraphics[width=0.28\textwidth]{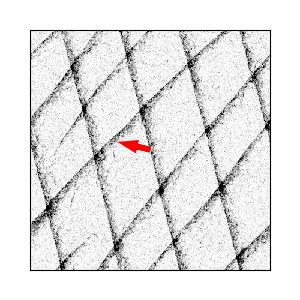}
 \includegraphics[width=0.28\textwidth]{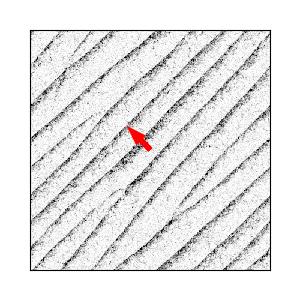}\\
 \includegraphics[width=0.28\textwidth]{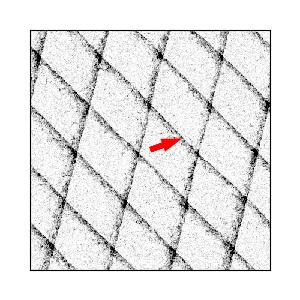}
 \includegraphics[width=0.28\textwidth]{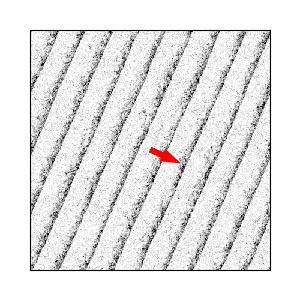}
 \includegraphics[width=0.28\textwidth]{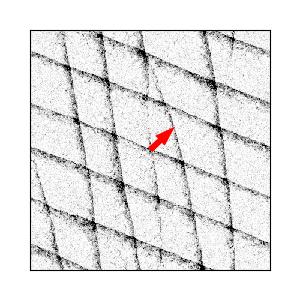}\\
\includegraphics[width=0.28\textwidth]{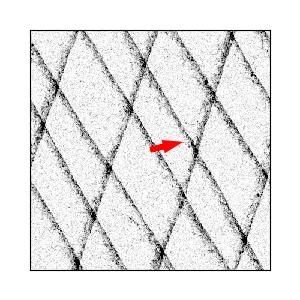}
\end{center}
\end{figure}

\newpage
$\eta=0.37$.
\begin{figure}[H]
\begin{center}
 \includegraphics[width=0.28\textwidth]{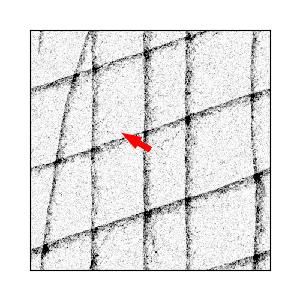}
 \includegraphics[width=0.28\textwidth]{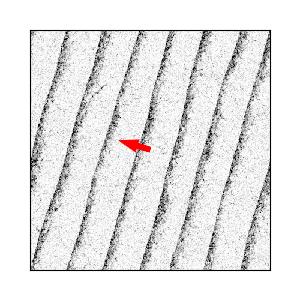}
 \includegraphics[width=0.28\textwidth]{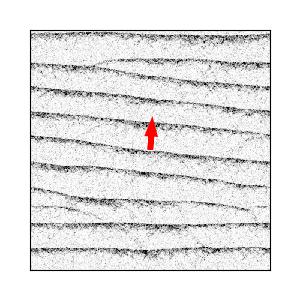}\\
 \includegraphics[width=0.28\textwidth]{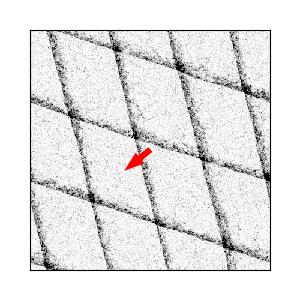}
 \includegraphics[width=0.28\textwidth]{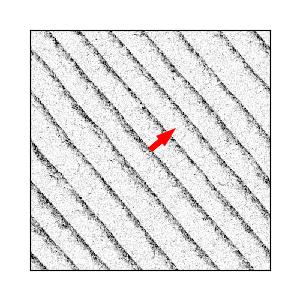}
 \includegraphics[width=0.28\textwidth]{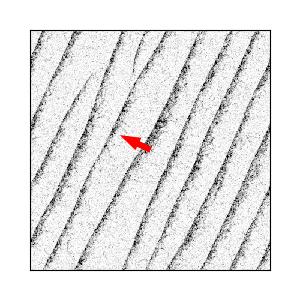}\\
 \includegraphics[width=0.28\textwidth]{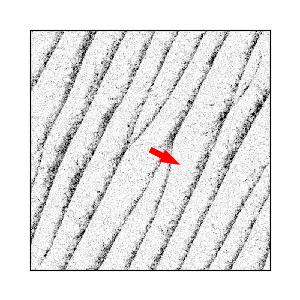}
 \includegraphics[width=0.28\textwidth]{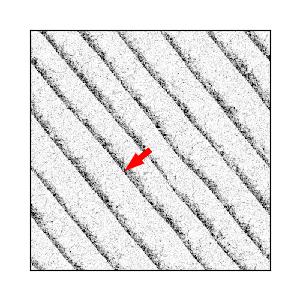}
 \includegraphics[width=0.28\textwidth]{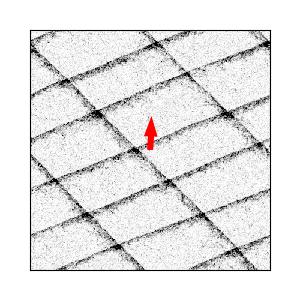}\\
\includegraphics[width=0.28\textwidth]{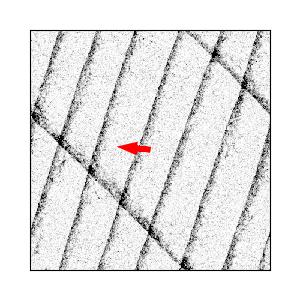}
\end{center}
\end{figure}

\newpage
$\eta=0.38$.
\begin{figure}[H]
\begin{center}
 \includegraphics[width=0.28\textwidth]{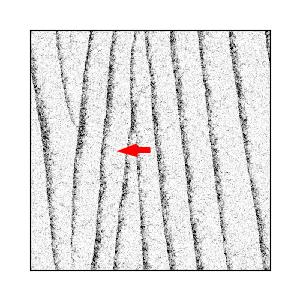}
 \includegraphics[width=0.28\textwidth]{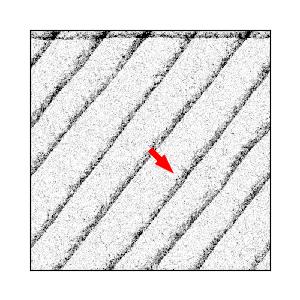}
 \includegraphics[width=0.28\textwidth]{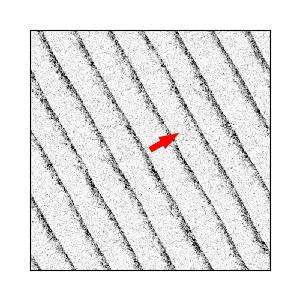}\\
 \includegraphics[width=0.28\textwidth]{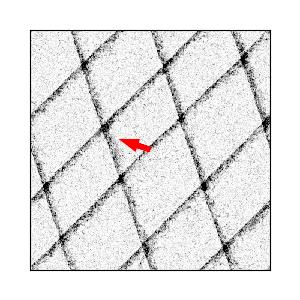}
 \includegraphics[width=0.28\textwidth]{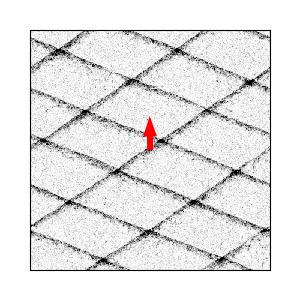}
 \includegraphics[width=0.28\textwidth]{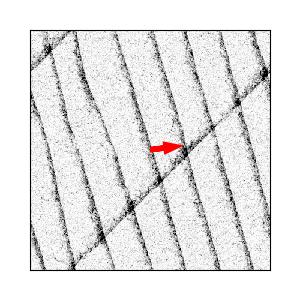}\\
 \includegraphics[width=0.28\textwidth]{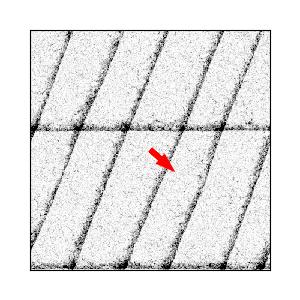}
 \includegraphics[width=0.28\textwidth]{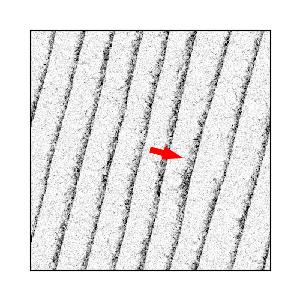}
 \includegraphics[width=0.28\textwidth]{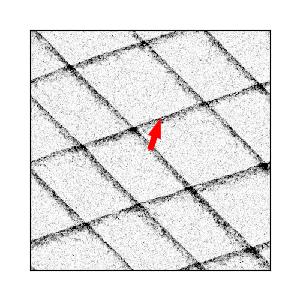}\\
\includegraphics[width=0.28\textwidth]{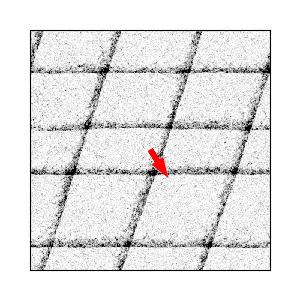}
\end{center}
\end{figure}

\newpage
$\eta=0.39$.
\begin{figure}[H]
\begin{center}
 \includegraphics[width=0.28\textwidth]{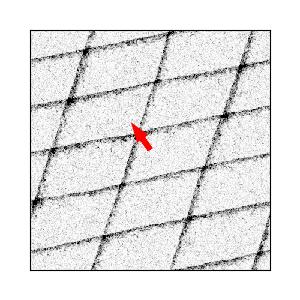}
 \includegraphics[width=0.28\textwidth]{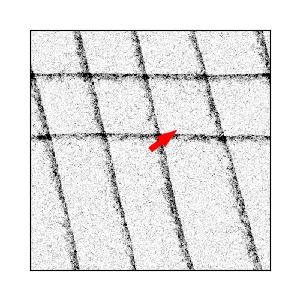}
 \includegraphics[width=0.28\textwidth]{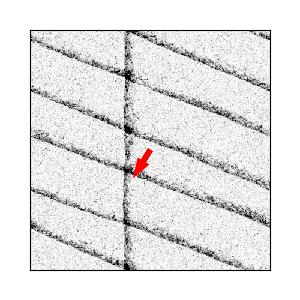}\\
 \includegraphics[width=0.28\textwidth]{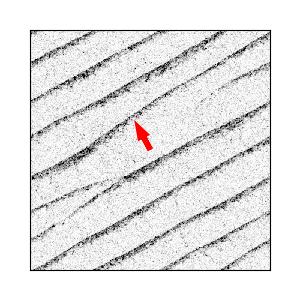}
 \includegraphics[width=0.28\textwidth]{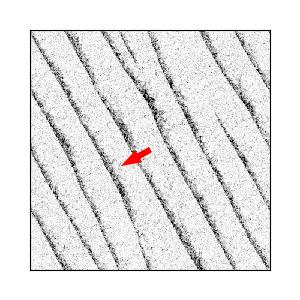}
 \includegraphics[width=0.28\textwidth]{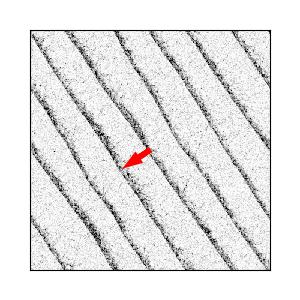}\\
 \includegraphics[width=0.28\textwidth]{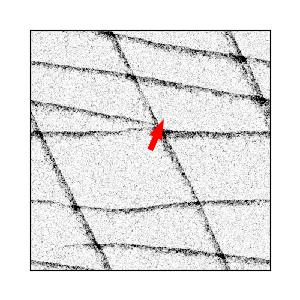}
 \includegraphics[width=0.28\textwidth]{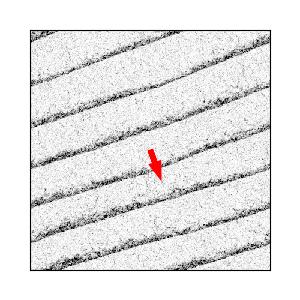}
 \includegraphics[width=0.28\textwidth]{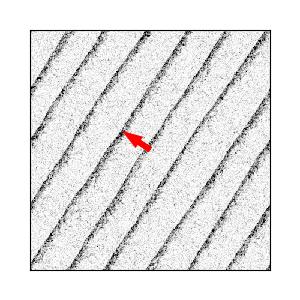}\\
\includegraphics[width=0.28\textwidth]{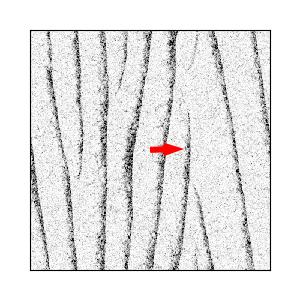}
\end{center}
\end{figure}

\newpage
$\eta=0.40$.
\begin{figure}[H]
\begin{center}
 \includegraphics[width=0.28\textwidth]{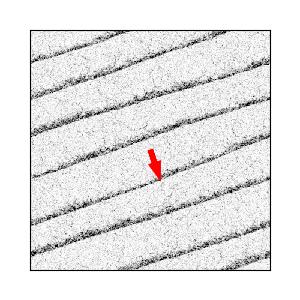}
 \includegraphics[width=0.28\textwidth]{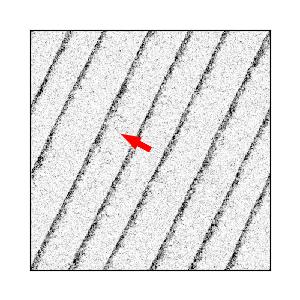}
 \includegraphics[width=0.28\textwidth]{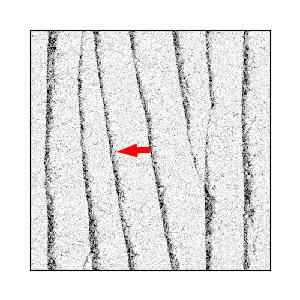}\\
 \includegraphics[width=0.28\textwidth]{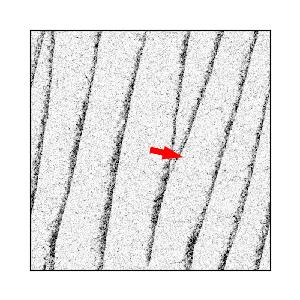}
 \includegraphics[width=0.28\textwidth]{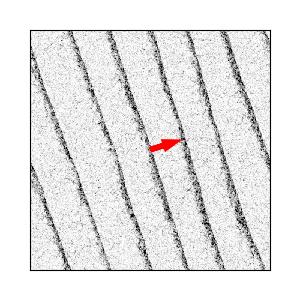}
 \includegraphics[width=0.28\textwidth]{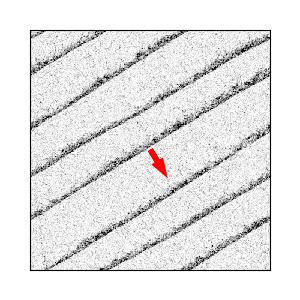}\\
 \includegraphics[width=0.28\textwidth]{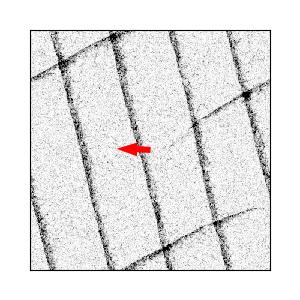}
 \includegraphics[width=0.28\textwidth]{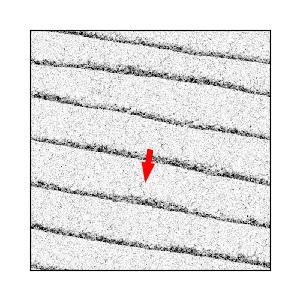}
 \includegraphics[width=0.28\textwidth]{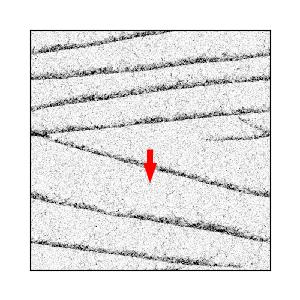}\\
\includegraphics[width=0.28\textwidth]{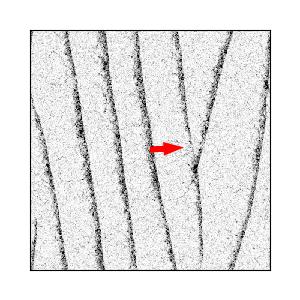}
\end{center}
\end{figure}

\newpage
$\eta=0.41$.
\begin{figure}[H]
\begin{center}
 \includegraphics[width=0.28\textwidth]{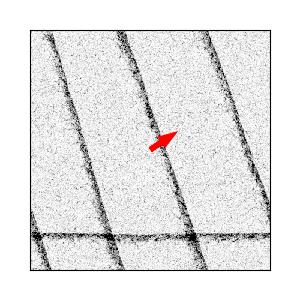}
 \includegraphics[width=0.28\textwidth]{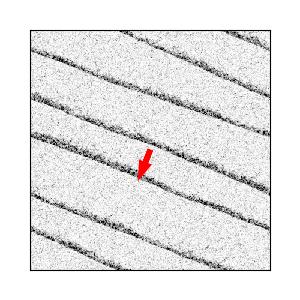}
 \includegraphics[width=0.28\textwidth]{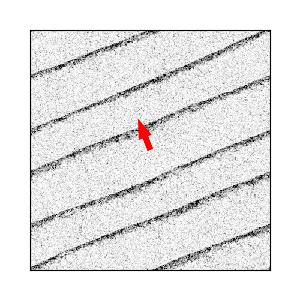}\\
 \includegraphics[width=0.28\textwidth]{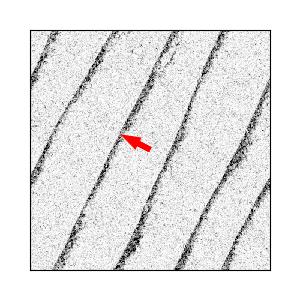}
 \includegraphics[width=0.28\textwidth]{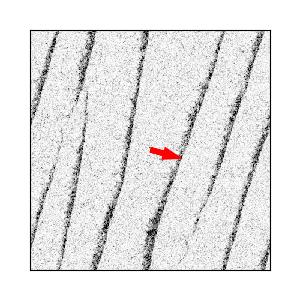}
 \includegraphics[width=0.28\textwidth]{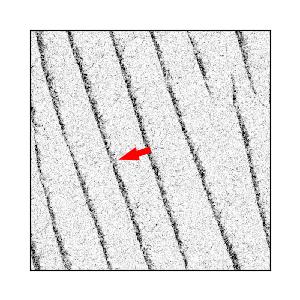}\\
 \includegraphics[width=0.28\textwidth]{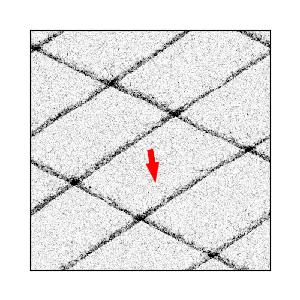}
 \includegraphics[width=0.28\textwidth]{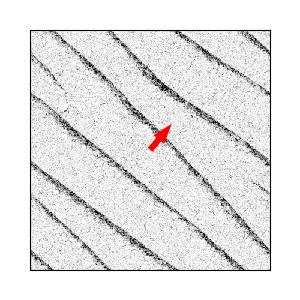}
 \includegraphics[width=0.28\textwidth]{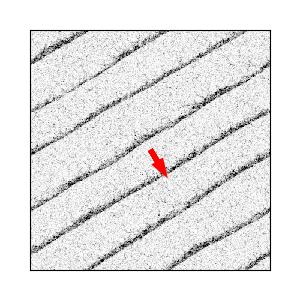}\\
\includegraphics[width=0.28\textwidth]{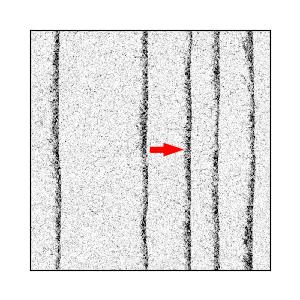}
\end{center}
\end{figure}

\newpage
$\eta=0.42$.
\begin{figure}[H]
\begin{center}
 \includegraphics[width=0.28\textwidth]{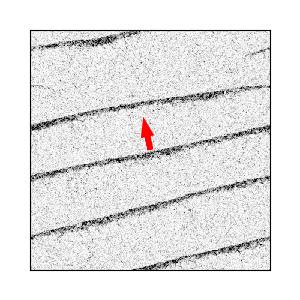}
 \includegraphics[width=0.28\textwidth]{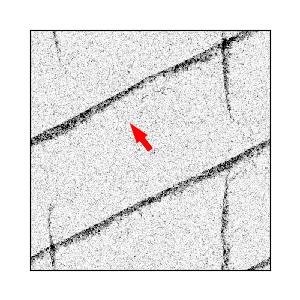}
 \includegraphics[width=0.28\textwidth]{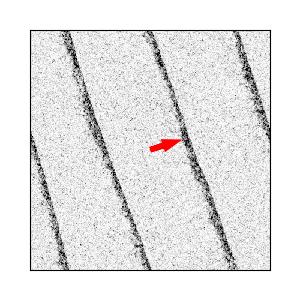}\\
 \includegraphics[width=0.28\textwidth]{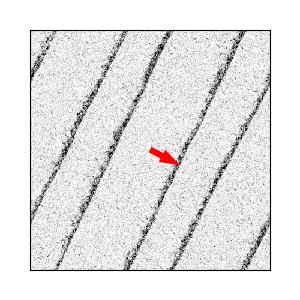}
 \includegraphics[width=0.28\textwidth]{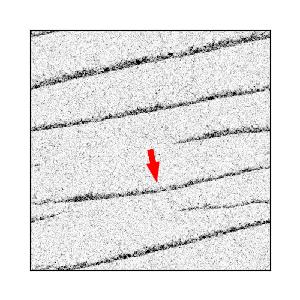}
 \includegraphics[width=0.28\textwidth]{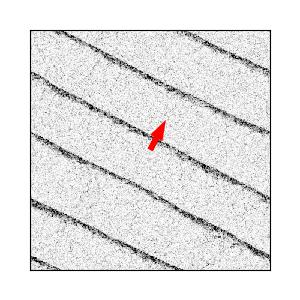}\\
 \includegraphics[width=0.28\textwidth]{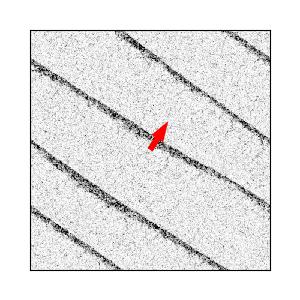}
 \includegraphics[width=0.28\textwidth]{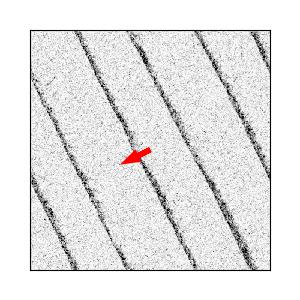}
 \includegraphics[width=0.28\textwidth]{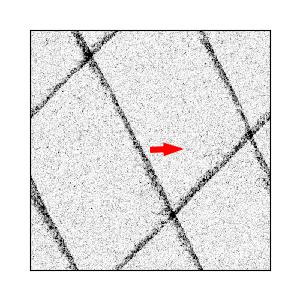}\\
\includegraphics[width=0.28\textwidth]{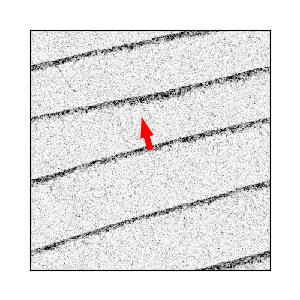}
\end{center}
\end{figure}

\newpage
$\eta=0.43$.
\begin{figure}[H]
\begin{center}
 \includegraphics[width=0.28\textwidth]{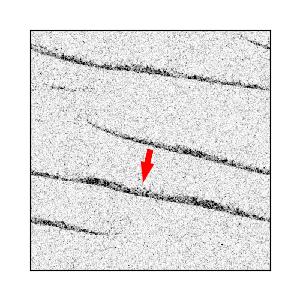}
 \includegraphics[width=0.28\textwidth]{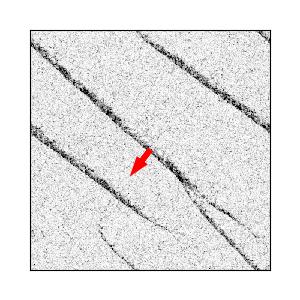}
 \includegraphics[width=0.28\textwidth]{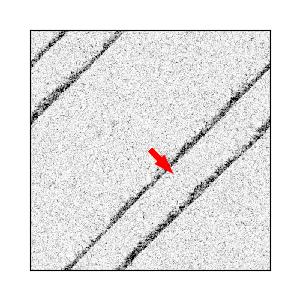}\\
 \includegraphics[width=0.28\textwidth]{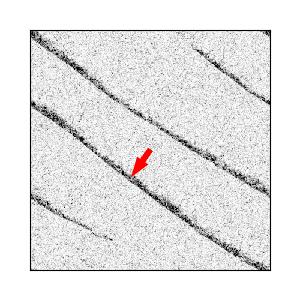}
 \includegraphics[width=0.28\textwidth]{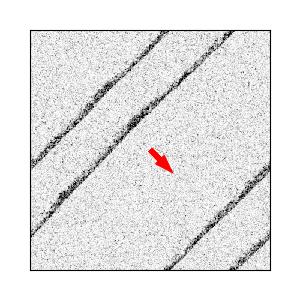}
 \includegraphics[width=0.28\textwidth]{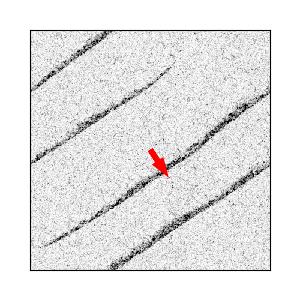}\\
 \includegraphics[width=0.28\textwidth]{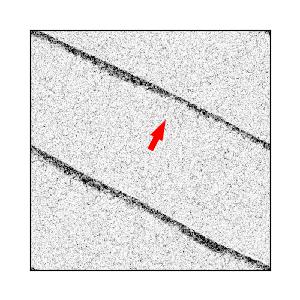}
 \includegraphics[width=0.28\textwidth]{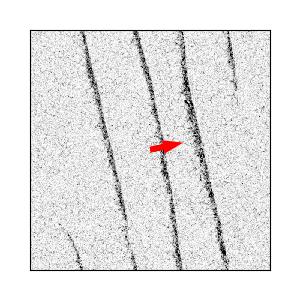}
 \includegraphics[width=0.28\textwidth]{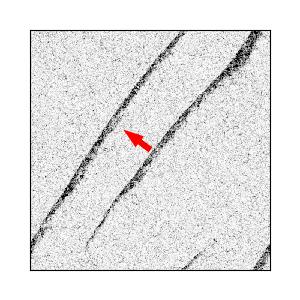}\\
\includegraphics[width=0.28\textwidth]{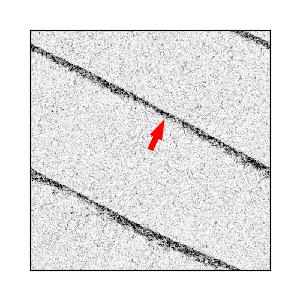}
\end{center}
\end{figure}

\newpage
$\eta=0.44$.
\begin{figure}[H]
\begin{center}
 \includegraphics[width=0.28\textwidth]{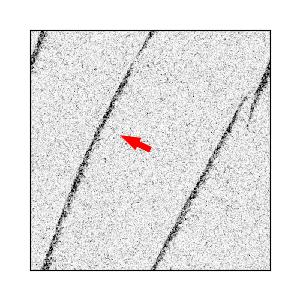}
 \includegraphics[width=0.28\textwidth]{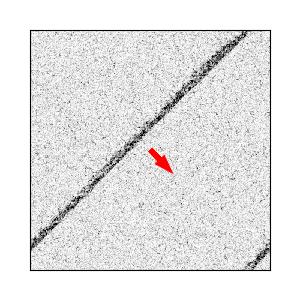}
 \includegraphics[width=0.28\textwidth]{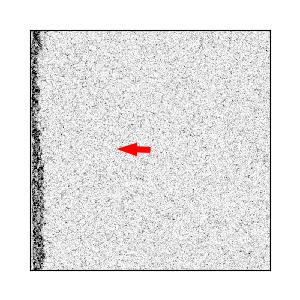}\\
 \includegraphics[width=0.28\textwidth]{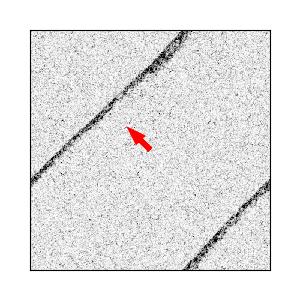}
 \includegraphics[width=0.28\textwidth]{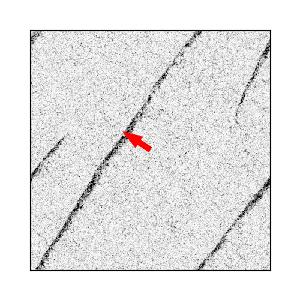}
 \includegraphics[width=0.28\textwidth]{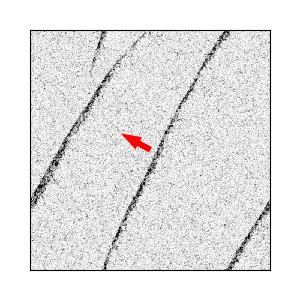}\\
 \includegraphics[width=0.28\textwidth]{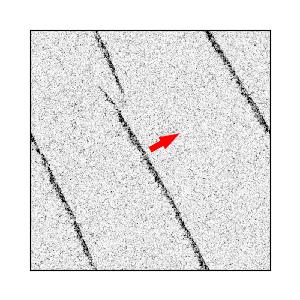}
 \includegraphics[width=0.28\textwidth]{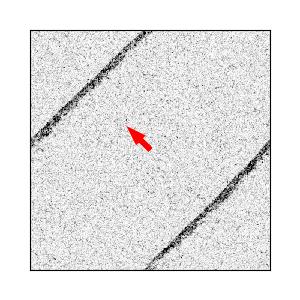}
 \includegraphics[width=0.28\textwidth]{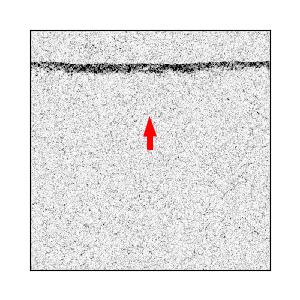}\\
\includegraphics[width=0.28\textwidth]{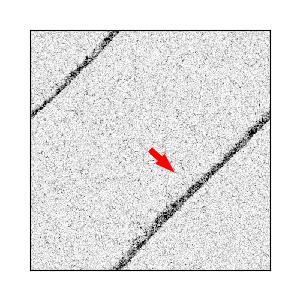}
\end{center}
\end{figure}

\newpage
$\eta=0.45$.
\begin{figure}[H]
\begin{center}
 \includegraphics[width=0.28\textwidth]{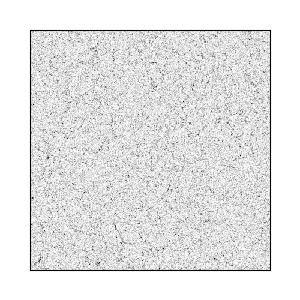}
 \includegraphics[width=0.28\textwidth]{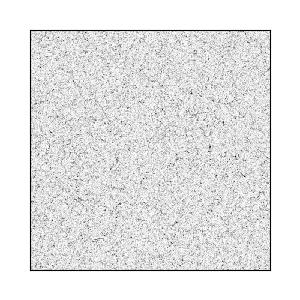}
 \includegraphics[width=0.28\textwidth]{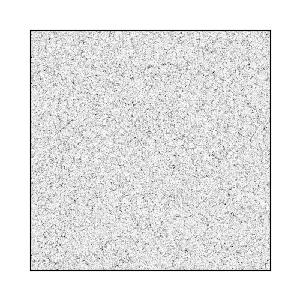}\\
 \includegraphics[width=0.28\textwidth]{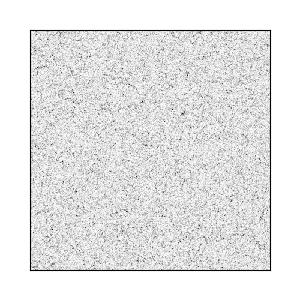}
 \includegraphics[width=0.28\textwidth]{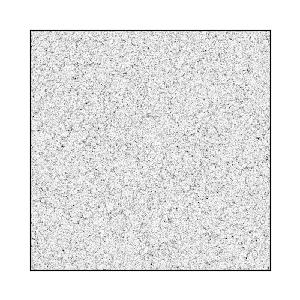}
 \includegraphics[width=0.28\textwidth]{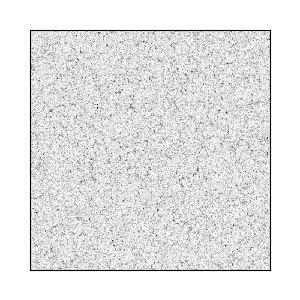}\\
 \includegraphics[width=0.28\textwidth]{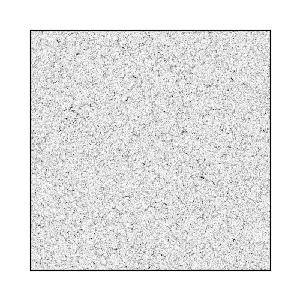}
 \includegraphics[width=0.28\textwidth]{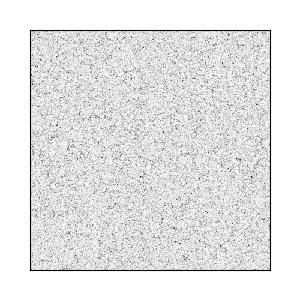}
 \includegraphics[width=0.28\textwidth]{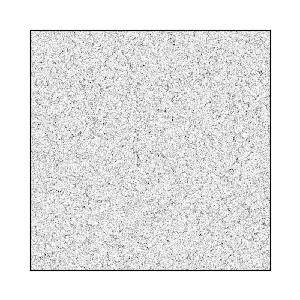}\\
\includegraphics[width=0.28\textwidth]{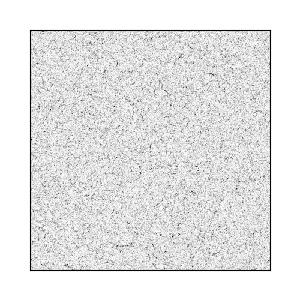}
\end{center}
\end{figure}
Snapshots for $\eta=0.46, 47, 48, 49$ look very similar to those at $\eta=0.45$ and are not presented here.

\end{document}